\documentclass[UTF8,a4paper]{article}
\usepackage{amsmath}
\usepackage{graphicx}
\usepackage{subfigure}
\usepackage{epstopdf}
\usepackage{geometry}
\usepackage{ulem}
\usepackage{indentfirst}
\usepackage{amsfonts,amssymb,dsfont}
\usepackage{setspace}
\usepackage{threeparttable}
\usepackage{amsmath,mathtools,amsthm}
\usepackage{array}
\usepackage{extarrows}
\usepackage{upgreek}

\makeatletter
\renewcommand*\env@matrix[1][\arraystretch]{%
  \edef\arraystretch{#1}%
  \hskip -\arraycolsep
  \let\@ifnextchar\new@ifnextchar
  \array{*\c@MaxMatrixCols c}}
\makeatother
\begin{document}


\title{\bf Interband and intraband transition, dynamical polarization and screening of the monolayer and bilayer silicene
in low-energy tight-binding model}
\author{Chen-Huan Wu
\thanks{chenhuanwu1@gmail.com}
\\Key Laboratory of Atomic $\&$ Molecular Physics and Functional Materials of Gansu Province,
\\College of Physics and Electronic Engineering, Northwest Normal University, Lanzhou 730070, China}

\maketitle
\vspace{-30pt}
\begin{abstract}
We investigate the interband and intraband transition of the monolayer and AB-stacked bilayer silicene in low-energy tight-binding model
under the electric field,
where we focus on the dynamical polarization function, screening due to the charged impurity, and the plasmon dispersion.
We obtain the logarithmically divergen polarization function within the random-phase-approximation (RPA) whose logarithmic singularities
corresponds to the discontinuities of the first derivative which is at the momentum ${\bf q}=2{\bf k}_{F}$ in static case and indicate
the topological phase transition point from the gapless semimetal to the gapped band insulator.
We also obtain the power-law-dependent Friedel oscillation which can be enhanced by increasing the Rashba-coupling,
that can contribute to the screened potential of the charged impurity which scale as $\sim r^{-1/2}$ in the short distance from the impurity
and scale as $\sim r^{-1/3}$ in the long distance from the impurity.
In the single-particle excitation regime with the electron-hole continuum, the interband and intraband transition happen,
and the plasmon dispersion, which we mainly focus on the optical plasmon (which $\sim\sqrt{{\bf q}}$ in long-wavelength limit) in this paper,
start to damped into the electron-hole pairs due to the nonzero imaginary part of the polarization function.
In low-frequency regime where the collective behavior and optical properties of the Dirac material 
relys more on the frequency than the fine structure constant,
the intraband transition is dominate and it's found that completely undamped in the static case ($\omega=0$),
which is due to the absence of the imaginary dynamic polarization.
We also observe the linear (weakly damped) plasmon model for the classical bilayer silicene which is similar to the high-energy $\pi$-plasmon
or the case of conducting substrate which with strong metallic screening in the bulk semiconductor.
For the large carrier density, we find the plasmon diapersion has $\omega_{p}\sim n^{1/2}$ which consistent with the quadratic dispersion around the Dirac-point
like the bilayer silicene 
with the effective mass about the interlayer hopping
(esperially when taking the Rashba-coupling and exchange field into consider) or the normal two-dimension electron gas,
while in the little concentration limit, $\omega_{p}\sim n^{1/4}$ which consistent with the linear dispersion like the monolayer silicene.
Under the nonmagnetic impurity scattering, the Thomas-Fermi decay and Friedel oscillation can easily be observed 
due to the strong spin-orbit couopling of the bilayer silicene even we don't take the Rashba-coupling into consider.

\begin{large}

\end{large}
\end{abstract}
\begin{large}
\section{Introduction}

Through the investigation of the electron transport properties of the monolayer and bilayer silicene as well
as there on-site Hubbard U-dependent phase transitions\cite{Wu C H},
we confirm that, the linear dispersion relation $|\varepsilon|=\hbar v_{F}|{\bf k}|$ near the Dirac-point,
(the Fermi velocity $v_{F}$ is treated as $5.5\times 10^{5}$ in this paper)
tends to quadratic dispersion for the AB-stacked bilayer silicene which with the finite density of state (DOS) 
and the screened long-range Coulomb scattering by the charged impurity.
The latter is common in the AB-stacked bilayer silicene or graphene and their multilayer bulk or nanoribbon form\cite{Ezawa M},
except that, the quadratic dispersion which is governed by the infrared divergence\cite{Girotti H O}
may diverges the susceptibility and logarithmically diverges the effective energy dispersion and the DOS in low-energy region
away from the linear dispersion under the effects of Coulomb coupling.
The resulting dispersion is $\varepsilon\sim \pm \hbar v_{F}|{\bf k}|(1+g_{0}{\rm ln}\frac{\Lambda_{0}}{\Lambda}g({\bf k}))^{-1}$
where $g_{0}=2\pi e^{2}/\epsilon_{0}\epsilon v_{F}$ is the dimensionless Coulomb coupling 
(effective fine structure constant) 
which scale to the zero here,
and $g({\bf k})=2\pi e^{2}/\epsilon_{0}\epsilon{\bf k}$ is the universe Coulomb coupling,
where the static background dielectric constant for the air/SiO$_{2}$ substrate is $\epsilon=2.45$ ($\epsilon_{SiO_{2}}=3.9$)
and $\epsilon_{0}$ is the dielectric constant of vacuum.
$\Lambda_{0}$ is the bare cutoff which $\sim t$ here ($t=1.6$ eV is the nearest neighbor hopping for a monolayer silicene) and
$\Lambda<\Lambda_{0}$ is in the low-energy range.
For the case of small bare Coulomb coupling $g_{0}\ll 1$, the renormalized Coulomb coupling and Fermi velocity show strongly 
frequency-dependence while the momentum-dependence is logarithmically decrese\cite{Wu C H,Aleiner I L}.
while for the case of $g\gg 1$ which away from the frequency-domain, they tends to momentum-dependent.
It's also found that the antiferromagnetic (AFM) or ferromagnetic (FM) excitonic instability in AB-bilayer silicene or graphene,
which with the gapless parabolic dispersion,
occur even under the strong screening of the long-range Coulomb scattering by the charged impurity\cite{Min H,Nandkishore R},
and with the interaction obeys the $1/r$ Hubbard model.
The magnetic instability may leads to the gapless band structure even for the triplet exciton\cite{Baskaran G}.

\section{Model}
In tight-binding model,
The Hamiltonian in low-energy Dirac theory are
given in
a non-Hermitian form\cite{Wu C H,Wu C H2,Liu C C,Ezawa M22,Ezawa M4,Ezawa M3,Ezawa M} 
\begin{equation} 
\begin{aligned}
H_{monolayer}=&t\sum_{\langle i,j\rangle ;\sigma}c^{\dag}_{i\sigma}c_{j\sigma}
+i\frac{\lambda_{{\rm SOC}}}{3\sqrt{3}}\sum_{\langle\langle i,j\rangle\rangle ;\sigma\sigma'}\upsilon_{ij}c^{\dag}_{i\sigma}\sigma^{z}_{\sigma\sigma'}c_{j\sigma'}
- i\frac{2R}{3}\sum_{\langle\langle i,j\rangle\rangle ;\sigma\sigma'}c^{\dag}_{i\sigma}(\mu \Delta({\bf k}_{ij})\times {\bf e}_{z})_{\sigma\sigma'}c_{i\sigma'}\\
&+iR_{2}(E_{\perp})\sum_{\langle i,j\rangle;\sigma\sigma'}c^{\dag}_{i\sigma}(\Delta({\bf k}_{ij})\times {\bf e}_{z})_{\sigma\sigma'}c_{i\sigma'}
-\frac{\overline{\Delta}}{2}\sum_{i\sigma} c^{\dag}_{i\sigma}\mu E_{\perp}c_{i\sigma}\\
&+M_{s}\sum_{i\sigma}c^{\dag}_{i\sigma}\sigma_{z}c_{i\sigma}
+M_{c}\sum_{i\sigma}c^{\dag}_{i\sigma}c_{i\sigma}
+U\sum_{i}\mu n_{i\uparrow}n_{i\downarrow},
\end{aligned}
\end{equation}
where $t=1.6$ eV is the nearest-neoghbor hopping which contains the contributions from both the $\pi$ band 
and $\sigma$ band.
The gap function is $\Delta({\bf k})={\bf d}({\bf k})\cdot{\pmb \sigma}$ which in a coordinate independent but spin-dependent representation.
The ${\bf k}$-dependent unit vector ${\bf d}({\bf k})$ here has
${\bf d}({\bf k})=[t'_{SOC}{\rm sin}k_{x},t'_{SOC}{\rm sin}k_{y},M_{z}-2B(2-{\rm cos}k_{x}+{\rm cos}k_{y})]$
for the BHZ model,
where $B$ is the BHZ model
-dependent parameter and $M_{z}$ the Zeeman field term which dominate the surface magnetization
but can be ignore when a strong electric field or magnetic field is applied.
$\langle i,j\rangle$ and $\langle\langle i,j\rangle\rangle $ denote the nearest-neighbor (NN) pairs and the next-nearest-neighbor (NNN) pairs, respectively. 
$\mu=\pm 1$ denote the $A$ ($B$) sublattices. 
Here 
${\bf d}({\bf k}_{ij})=\frac{{\bf d}_{ij}}{|{\bf d}{ij}|}$ is the NNN hopping vector.
$\lambda_{SOC}=3.9$ meV is the intrinsic spin-orbit coupling (SOC) strength which is much larger than the monolayer graphene's (0.0065 meV\cite{Guinea F}).
$R$ is the small instrinct Rashba-coupling due to the low-buckled structure, which is related to the helical bands
(helical edge states) and the SDW in silicene, and it's disappear in the Dirac-point ($k_{x}=k_{y}=0$).
$R_{2}(E_{\perp})$ is the extrinsic Rashba-coupling induced by the electric field.
The existence of $R$ breaks U(1) spin conservation (thus the $s^{z}$ is no more conserved) and the mirror symmetry of silicene lattice.
$M=M_{s}+M_{c}$ is the exchange field which breaks the spatial-inverse-symmetry and the $M_{s}$ is related to the out-of-plane FM exchange field
with parallel alignment of exchange magnetization
and $M_{c}$ is related to the CDW,
which endows sublattice pseudospin the $z$-component\cite{Ezawa M6}.
While for the out-of-plane AFM exchange field $M_{s}^{AFM}$ which is not contained here with antiparallel alignment of exchange magnetization.
Here the $M$ is applied perpendicular to the silicene, and it can be rised by proximity coupling to the ferromagnet\cite{Ezawa M}.
Thus the induced exchange magnetization along the $z$-axis between two sublattices-plane is
related to the SOC, Rashba-coupling, and even the Zeeman-field since it will affects the magnetic-order in $z$-direction.
In fact, if without the exchange field and only exist the SOC, the spin-up and spin-down states won't be degenerates but will mixed around the crossing
points between the lowest conduction band and the highest valence band just like the spin-valley-polarized semimetal (SVPSM).
Note that here we follow the definition of semimetal that the conduction band and valence band have a small overlap,
no matter the two bands are with linear dispersion in the crossing point or 
parabolic dispersion (quadratic) in the crossing point like the Fermi point of the AB-stacked bilayer silicene or graphene.
$\upsilon_{ij}=({\bf d}_{i}\times{\bf d}_{j})/|{\bf d}_{i}\times{\bf d}_{j}|=1(-1)$ 
when the next-nearest-neighboring hopping of electron is toward left (right),
with ${\bf d}_{i}\times{\bf d}_{j}=\sqrt{3}/2(-\sqrt{3}/2)$.
The term contains the exchange field $M$ is the staggered potential term induced by the buckled structure
which breaks the particle-hole symmetry.
Here the coordinate-independent representation of the Rashba-coupling terms is due to the broken of inversion symmetry as well as the mirror symmetry.
The last term is the Hubbard term with on-site interaction $U$ which doesn't affects the bulk gap here but affects the edge gap.
Thus the $U$ is setted as zero within the bulk but nonzero in the edge, 
which is also consistent with the STM-result of silicene that the edge states have 
higher electron-density than the bulk.
And here we take account the on-site Hubbard interaction only and ignore the long-range ones which are screened by the finite DOS with high energy,
like the NN or NNN Coulomb repulsion, interlayer Coulomb repulsion,
and even the one with a range much larger that $a$ (like the Bohr radius in semiconductor).
There are two kinds of AB-stacked bilayer silicene: one
with the nearest layer distance as $d=$2.53 \AA\ and intra-layer bond length 2.32 \AA\ with the bulked distance $\overline{\Delta}=0.64$ \AA\ 
the smae as the monolayer one
and the another one with the nearest layer distance as $d=$2.92 \AA\ and intra-layer bond length 2.32 \AA\ with the lattice constant $a=3.88$
and the buckled distance $\overline{\Delta}=0.64$ \AA\
as plotted in the Fig.1.
Thus for the bilayer silicene, the eight-band tight-binding (TB) model in low-energy Dirac theory is
\begin{equation} 
\begin{aligned}
H_{bilayer}=&t\sum_{\langle i,j\rangle ,\sigma,l}c^{\dag}_{i\sigma l}c_{j\sigma l}
+i\frac{\lambda_{{\rm SOC}}}{3\sqrt{3}}\sum_{\langle\langle i,j\rangle\rangle ;\sigma\sigma'}\upsilon_{ij}c^{\dag}_{i\sigma l}\sigma^{z}_{\sigma\sigma'}c_{j\sigma' l}
- i\frac{2R}{3}\sum_{\langle\langle i,j\rangle\rangle ,\sigma\sigma',l}c^{\dag}_{i\sigma l}(\mu \Delta({\bf k}_{ij})\times {\bf e}_{z})_{\sigma\sigma' }c_{i\sigma' l}\\
&+iR_{2}(E_{\perp})\sum_{\langle i,j\rangle,\sigma\sigma',l}c^{\dag}_{i\sigma}(\Delta({\bf k}_{ij})\times {\bf e}_{z})_{\sigma\sigma'}c_{i\sigma'l}
-\frac{\overline{\Delta}}{2}\sum_{i\sigma l} c^{\dag}_{i\sigma l}\mu E_{\perp}c_{i\sigma l}\\
&+M_{s}\sum_{i\sigma l}c^{\dag}_{i\sigma l}\sigma_{z}c_{i\sigma l}
+M_{c}\sum_{i\sigma l}c^{\dag}_{i\sigma l}c_{i\sigma l}
+U\sum_{i,l}\mu n_{i,l\uparrow}n_{i,l\downarrow}+t_{1}\sum_{i,\sigma,l}c_{i}^{\dag}c_{j}\\
&+i\lambda_{SOC}^{{\rm int}}\sum_{i\in A_{1},j\in A_{2},\sigma}c^{\dag}_{i\sigma}(\mu \Delta({\bf k}_{ij})\times {\bf e}_{z})_{\sigma\sigma'}c_{i\sigma'}\\
&+i\lambda_{SOC}^{{\rm int}}\sum_{i\in B_{1},j\in B_{2},\sigma}c^{\dag}_{i\sigma}(\mu \Delta({\bf k}_{ij})\times {\bf e}_{z})_{\sigma\sigma'}c_{i\sigma'}\\
&+\left\{
\begin{array}{rcl}
t_{1}\sum_{i\in B_{1},j\in A_{2},\sigma} c^{\dag}_{i\sigma}\mu c_{j\sigma},&\ for\ 1st\ AB-stacked\ bilayer\ silicene,\\
t_{2}\sum_{i\in B_{1},j\in B_{2},\sigma} c^{\dag}_{i\sigma}\mu c_{j\sigma},&\ for\ 2nd\ AB-stacked\ bilayer\ silicene,
\end{array}
\right.
\end{aligned}
\end{equation}
where $l=\pm 1$ is the layer index, and $\lambda_{SOC}^{{\rm int}}=0.5$ meV is the interlayer SOC\cite{Ezawa M2},
$A_{1}$ belong to the upper layer and $A_{2}$ belong to the bottom one.
$t_{1}=2.025$ eV is the NN interlayer hopping\cite{Liu F} which is much larger than the van der Waals interaction.

In simplify, the Hamiltonian can be represented as $H={\pmb \tau}\cdot{\bf d}$,
where ${\pmb \tau}$ here describe the sublattice degrees-of-freedom which also brings the mass term,
while the valley degrees-of-freedom is contained in ${\bf k}$.
The $z$-component of three-dimension vector ${\bf k}$ is parallel to the orbital angular momentum $\hbar\hat{L}$,
due to the nuclear dipolar which is important 
for the gapless excitation,
the local spin density $\hat{I}$ for this model has
\begin{equation} 
\begin{aligned}
\hat{I}_{x}=&\frac{1}{2}(\psi^{\dag}_{\uparrow}\psi_{\downarrow}-\psi^{\dag}_{\downarrow}\psi_{\uparrow}),\\
\hat{I}_{y}=&\frac{1}{2}(\psi^{\dag}_{\uparrow}\psi_{\downarrow}+\psi^{\dag}_{\downarrow}\psi_{\uparrow}),\\
\hat{I}_{z}=&-i\psi_{\uparrow}\psi_{\uparrow},
\end{aligned}
\end{equation}
thus in unepitaxial case, the nonmetallic surface state is possible when the local perturbation coup to the $\hat{I}_{z}$
(i.e., the component of local spin density which is normal to the surface [100]).
Here such perturbation here may caused by the external magnetic field or the internal spin interaction,
in fact, for the thermodynamic quantitys in our tight-binding model, including the interband interaction and the orbital or spin susceptibility, etc.,
their time evolution is associate with these perturbations which may induce the quench effect as well as the band energy spectrum.
For the gapless low-energy tight-binding model, 
the charge and spin susceptibility obtained by the random-phase-approximation (RPA) are associated with the on-site Hubbard repulsion,
and are decrease and increase with the increasing on-ite Hubbard repulsion, respectively\cite{Liu F}.
They are also sensitive to the charge-density-wave (CDW) and spin-density-wave (SDW), respectively,
and their properties as well as the temperature-dependence can be well studied by the 
nuclear magnetic resonance (NMR),
and the inelastic neutron scattering.
That's different from the orbital susceptibility which is diamagnetic (negative) and anisotropic
as a result of the competition between the spin-up and spin-down carriers, 
and the diamagnetic momentum is larger that paramagnetic one.
In fact,
both the diamagnetic and paramagnetic response which with opposite magnetic moment (i.e.,
diamagnetic moment and paramagnetic moment with the spin carriers along the edge direction
carriers the up- and down- spin, respectively) are coexist in the silicene due to the interactions
between the magnetic field and the charge carriers with spin-up and spin-down, respectively,
and they are both increse with the temperature.
The Dirac-mass-dependent diamagnetic susceptibility at low-temperature is
\begin{equation} 
\begin{aligned}
\chi(\beta,m_{D})=\frac{-4e^{2}\hbar^{2}v_{F}^{2}}{6\pi c^{2}}\frac{1}{2|m_{D}|}{\rm tanh}(2m_{D}\beta),\\
\chi_{T\rightarrow 0}(m_{D})=\frac{-4e^{2}\hbar^{2}v_{F}^{2}}{6\pi c^{2}}\frac{1}{2|m_{D}|}\Theta(|m_{D}|-|\varepsilon|).
\end{aligned}
\end{equation}
where $\beta$ is the inverse temperature ($k_{B}=1$)
and the Dirac-mass here is $m_{D}=\eta\lambda_{{\rm SOC}}s_{z}-\frac{\overline{\Delta}}{2}E_{\perp}+Ms_{z})$
where we ignore the effect of the intrinsic and external Rashba-coupling.

The band structures are presented in the Fig.2,
where we carry out the first-principle (FP) density functional theory (DFT) calculations using
the QUANTUM ESPRESSO package\cite{Paolo} with the generalized gradient approximation (GGA).
We found that the bilayer silicene is no more exhibits the linear Dirac dispersion in the low-energy regime near the Dirac-point:
for 1st AB-stacked bilayer silicene, there is a overlap of 320 meV between the highest valence band and lowest conduction band, 
while for the 2nd AB-stacked bilayer silicene, the band crossing point is vanish.

\section{interband transmission and polarization}

We have deduced the low-temperture longitudunal in-plane conductivity (diagonal) as
\begin{equation} 
\begin{aligned}
\sigma_{xx}=\sigma_{yy}=\frac{\beta e^{2}}{S}\sum_{m}f_{m}(1-f_{m})\frac{\langle m|v_{x}|m\rangle\langle m|v_{y}|m\rangle}{\omega+i m_{D}+2\Gamma}
\end{aligned}
\end{equation}
where $S=3\sqrt{3}a^{2}/2$ is the area of unit cell (Wigner-Seitz cell), 
$\beta$ is the inversed temperature,
$\omega=(2n+1)\pi/\beta$ is the fermionic Matsubara frequency where $\beta$ is the inverse temperature.
$v_{x}=\frac{\partial}{\hbar\partial k_{x}}$ is the velocity operator.
The longitudunal conductivity is related to the interband transmission,
and the screened Coulomb scattering by the charged impurity with the transferd cyclotron orbit if under the magnetic field
with
the cyclotron resonance frequency $\omega_{c}=\frac{\sqrt{2}\hbar v_{F}}{\ell_{B}}=\frac{|eB|}{cm^{*}}$
where $\ell_{B}=\sqrt{\hbar c/|eB|}$ is the magnetic length which play the role of quantized cyclotron orbit radius
in lowest Landau level (LLL) (n=0) $R_{0}=\ell_{B}$ and the quantized cyclotron orbit radius for $n\neq 0$ is $R_{n}=\sqrt{2n}\ell_{B}$.
In this case, the kinetic energy of a single-electron is $\sim \hbar\omega_{c}$\cite{Kotov V N}.
It's also found that, with the increase of chemical potential,
the spectral weight of intraband transition is rised for the real part of longitudunal in-plane conductivity $\sigma_{xx}$\cite{Wu C H,Tabert C J}

For the electron-hole pair within the process of interband transition,
the scattering matrix can be consisted by the two pairs: transmission (including the normal scattering (specular one or the backscattering) 
and Andreev one with a s-wave superconductor) and 
reflection (including the specular scattering and Andreev one) of the electrons,
and the scatterings are odd parity for the particle-hole transformation,
e.g., $|h{\bf k}\rangle=e^{2i\phi_{k}}|e{\bf k}\rangle$,
where $|h{\bf k}\rangle$ and $|e{\bf k}\rangle$ are the electron state and hole state, respectively,
and $e^{2i\phi_{k}}$ is the pseudospin(valley)-dependent odd parity scattering factor
(which is easy to proved by carry out the particle-hole transition as $c_{i\uparrow}\rightarrow c_{i\uparrow},c_{i\downarrow}\rightarrow (-1)^{i}c_{i\downarrow}$
in a AFM ordered spin pattern. see the below text).
We can represent it in the single-terminal travelling model as
\begin{equation} 
\begin{aligned}
\begin{pmatrix}[1.5] |h{\bf k}\rangle\\
|h{\bf k}\rangle^{\dag}\end{pmatrix}
=
\begin{pmatrix}[1.5] 
0&-1\\
-1&0
\end{pmatrix}
\begin{pmatrix}[1.5] 
|e{\bf k}\rangle\\
|e{\bf k}\rangle^{\dag}
\end{pmatrix}
=
\begin{pmatrix}[1.5] 
|-e{\bf k}\rangle^{\dag}\\
|-e{\bf k}\rangle
\end{pmatrix}
\end{aligned}
\end{equation}
or for the four-terminal one,
\begin{equation} 
\begin{aligned}
\begin{pmatrix}[1.5] |h_{1}{\bf k}\rangle\\
|h_{1}{\bf k}\rangle^{\dag}\\
|h_{2}{\bf k}\rangle\\
|h_{2}{\bf k}\rangle^{\dag}\end{pmatrix}
=
\begin{pmatrix}[1.5] 
0&-1&0&0\\
-1&0&0&0\\
0&0&0&1\\
0&0&1&0
\end{pmatrix}
\begin{pmatrix}[1.5] 
|e_{1}{\bf k}\rangle\\
|e_{1}{\bf k}\rangle^{\dag}\\
|e_{2}{\bf k}\rangle\\
|e_{2}{\bf k}\rangle^{\dag}
\end{pmatrix}
=
\begin{pmatrix}[1.5] 
|-e_{1}{\bf k}\rangle^{\dag}\\
|-e_{1}{\bf k}\rangle\\
|e_{2}{\bf k}\rangle^{\dag}\\
|e_{2}{\bf k}\rangle
\end{pmatrix},
\end{aligned}
\end{equation}
and the quantized charge conductance can be obtained by $\sigma_{xy}=e^{2}/(2h)$ in the Landauer-B\"{u}ttiker framework.
The equation of motion of the time-dependent electron/hole occupation $n_{{\bf k}}(t)$ can be obtained by the Boltzmann function
as
(we set the charge of electron $e=1$)
\begin{equation} 
\begin{aligned}
\frac{d}{dt}n_{{\bf k}}(t)=-\frac{2\alpha_{{\bf k}}(t)}{\hbar}E_{\perp}(t){\rm Im}[\Pi({\bf k},\omega)]+E_{\perp}(t)\frac{\partial}{\partial {\bf k}}n_{{\bf k}}(t),
\end{aligned}
\end{equation}
where $\alpha_{{\bf k}}(t)$ is the interband Coulomb dipole matrix elements which is real and time-dependent
which can leads to the high-harmonic generation (HHG) as a result of the dipole radiation and then
$E_{\perp}(t)=E_{0}{\rm sin}(\omega (t+t_{0}))$ is the time-dependent electric field of the laser pulse which perpendicular to the silicene plane.
and the resulting electric field force is ${\bf F}=-E_{\perp}(t){\bf e}_{z}$.
The lattice Green's function in helicity basis $G_{{\bf k}}(E_{m}-E_{n})=[E_{m}-E_{n}-(\hbar\omega+2i\Gamma)-\mu]^{-1}$
which can be obtained by the retarded form 
analytical continuation as $i\hbar\omega_{l}\rightarrow \hbar(\omega_{l}+i m_{D})$\cite{Wu C H} where $ m_{D}=0^{+}$ is a small positive quantity
and it has $\omega+i m_{D}\rightarrow 0$ in dc-limit.
The scattering rate $\Gamma$ due to the charged impurity (or the Dirac quasiparticles) here is defined as
\begin{equation} 
\begin{aligned}
\Gamma=\frac{1}{2\tau}=\frac{\pi n}{\hbar}V^{2}.
\end{aligned}
\end{equation}
Here the charged impurity density $n$ is momentum-independent for the single-impurity case.
The $\Gamma$ can be estimated as 0.01$t=0.016$ eV here and note that the effect of SOC is ignored in this scattering process
due to the large chemical potential.
Under the magnetic field, the scattering rate also represented teh width of the Landau level.

In the analytical continuation, the free-particle polarization function (the dynamical susceptibility) which related to the current-current correlation function,
can be obtained as
\begin{equation} 
\begin{aligned}
\Pi({\bf q},i\Omega)=-\frac{4e^{2}}{\beta}\int\frac{d^{2}k}{4\pi^{2}}{\rm Tr}[v_{\alpha}G_{{\bf k}}(i\omega+\Omega+i m_{D})v_{\beta}G_{{\bf k}}(i\omega)],
\end{aligned}
\end{equation}
in bubble diagram where the vertex function is not considered.
The $\Omega$ is the bosonic Matsubara frequency (like the photon) which $\Omega=2\pi m/\beta$ with $m=0,\pm 1,\pm 2,\cdot\cdot\cdot$.
Here the spectral-represented Grenn's function $G_{{\bf k}}(i\omega)=\int^{\infty}_{-\infty}\frac{d\omega}{2\pi}\frac{A(\omega,{\bf k})}{i\omega+\mu-\omega}$
with $A(\omega,{\bf k})$ the spectral weight.
$v_{\alpha}$ and $v_{\beta}$ denote the two velocity operators with the leads $\alpha,\beta=x,y,z$,
which
\begin{equation} 
\begin{aligned}
v_{x}=v_{F}I\gamma_{x},\ v_{y}=v_{F}I\gamma_{y},\ v_{z}=v_{F}I\gamma_{z},
\end{aligned}
\end{equation}
with $I$ the $4\times 4$ identity matrix,
and the $4\times 4$ Gamma matrices: $\gamma_{x}=\sigma_{z}\otimes i\sigma_{y}$, 
$\gamma_{y}=\sigma_{z}\otimes i\sigma_{x}$, $\gamma_{z}=i\sigma_{z}\otimes i\sigma_{z}$.
Or in the current-current correlated form where real frequency $\Omega$ is used here
\begin{equation} 
\begin{aligned}
\Pi({\bf q},i\Omega)=\frac{1}{S}\int^{\beta}_{0}dt\langle\mathcal{T}J_{\alpha}(\tau)J_{\beta}(0)\rangle e^{i\omega \tau}
\end{aligned}
\end{equation}
where $S$ is the sample area of a unit cell which is setted as 3$\sqrt{3}a^{2}/2$ in this paper, 
$\mathcal{T}$ is the time ordering operator and we assume the spatial-homogeneous and thus neglect the spatial dependence here
as $J(\tau)=\sum_{r}j(\tau,r)$,
thus the current operator is coordinate-independent,
which is different from the spatial-quantum-critical-point problem in two-dimension quantum system\cite{Sachdev S},
and note that in the following, the $n_{{\bf k}}(t)$ is evolution with the real time by the retarded form analytically continuing.

\section{static polarization in four band model}

For the static polarization in the case of $\mu< m_{D}$,
which is pure real due to the absence of imaginary frequency $i\omega=0$ and thus ${\rm Im}$,
and leading to $\Pi_{0}({\bf k})=\Pi({\bf k},i\omega=0)$,
with the strong momentum-dependence due to the strong Coulomb coupling, 
and its largest eigenvalue is the static homogeneous polarization which will becomes more homogeneous due to the neglect of some unimportant orbitals,
and the eigenvector which corresponding to the largest eigenvalue determines the dominant spin fluctuations\cite{Zhang L D}.
At zero temperature, the Pauli susceptibility which is proportional to the total density of states (DOS) at the Fermi surface is
$\chi(0)=\chi_{a_{2}a_{2}}^{(0)a_{1}a_{1}}(k,0)=\sum_{ab}n_{ab}(0)$ where $n_{ab}$ is the single-spin DOS at the Fermi surface for the bands in cell $a$ and $b$.
When such Hubbard interaction $U\neq 0$,
the charge and spin renormalized
 susceptibilities which enhanced by RPA are\cite{Liu F}
\begin{equation} 
\begin{aligned}
\chi^{(s)}(k,i\omega_{M})=[I-\chi^{(0)}(k,i\omega_{M}){\bf U}^{(s)}]^{-1}\chi^{(0)}(k,i\omega_{M}),\\
\chi^{(c)}(k,i\omega_{M})=[I+\chi^{(0)}(k,i\omega_{M}){\bf U}^{(c)}]^{-1}\chi^{(0)}(k,i\omega_{M}),
\end{aligned}
\end{equation}
where $I$ is the identity matrix, ${\bf U}$ is the $16\times 16$ matrix and there are only 40 nonzero elements for the spin 
susceptibility
and 28 nonzero elements for the charge 
susceptibility:
\begin{equation} 
\begin{aligned}
{\bf U}^{(s)a_{m}a_{m}}_{a_{m}a_{m}}=&U,\ {\bf U}^{(s)a_{m}a_{m}}_{a_{n}a_{n}}=\frac{J}{2},\ {\bf U}^{(s)a_{m}a_{n}}_{a_{m}a_{n}}=\frac{J}{4},\ {\bf U}^{(s)a_{n}a_{m}}_{a_{m}a_{n}}=t_{p},\\
{\bf U}^{(c)a_{m}a_{m}}_{a_{m}a_{m}}=&U,\ {\bf U}^{(c)a_{m}a_{n}}_{a_{m}a_{n}}=\frac{3J}{4},\ {\bf U}^{(c)a_{m}a_{n}}_{a_{n}a_{m}}=t_{p}.\\
(m,n=&1,2,3,4)
\end{aligned}
\end{equation}
We see that except the four intraband elements ${\bf U}^{(s)a_{m}a_{m}}_{a_{m}a_{m}}$, there are also some nonzero off-diagonal elements (interband) which is 
the result of considering the Hund's rule coupling.
The charge and spin susceptibility matrices here are also $16\times 16$.
Through the charge/spin fluctuations (or the charge/spin susceptibility),
the pairing scattering between the cooper pairs between different cells through the spin or charge fluctuations, 
i.e., $(ka_{1},-ka_{2})\rightarrow (k'a_{3},-k'a_{4})$ which also scatter to a new FS sheet,
is govern by the interaction-Hamiltonian in the momentum-space
\begin{equation} 
\begin{aligned}
H_{{\rm int}}=\sum_{a_{1}a_{2}a_{3}a_{4},\sigma\sigma',kk'}\frac{t_{p}}{2}c^{\dag}_{a_{1}\sigma}(k)c^{\dag}_{a_{2}\sigma' }(-k)c_{a_{3}\sigma}(-k')c_{a_{4}\sigma' }(k'),
\end{aligned}
\end{equation}
where $t_{p}$ is the pair-hopping,
give rise to the effective interaction in RPA level
\begin{equation} 
\begin{aligned}
U_{{\rm eff}}=\frac{1}{N}\sum_{a_{1}a_{2}a_{3}a_{4},kk'}\Gamma^{a_{1}a_{2}}_{a_{3}a_{4}}(k,k',\omega)c^{\dag}_{a_{1}}(k)c^{\dag}_{a_{2}}(-k)c_{a_{3}}(-k')c_{a_{4}}(k'),
\end{aligned}
\end{equation}
with effective pairing interaction vertex from the generalized RPA $\Gamma^{a_{1}a_{2}}_{a_{3}a_{4}}(k,k',\omega)$ in
spin-singlet and spin-triplet representations in momentum-space are
\begin{equation} 
\begin{aligned}
\Gamma^{(s)a_{1}a_{2}}_{a_{3}a_{4}}(k,k',\omega)=&\left[\frac{3}{2}{\bf U}^{(s)}\chi^{(s)}(k-k',\omega)\chi^{(s)}-\frac{1}{2}{\bf U}^{(c)}\chi^{(c)}(k-k',\omega)\chi^{(c)}
+\frac{1}{2}{\bf U}^{(s)}+\frac{1}{2}{\bf U}^{(c)}\right]^{a_{1}a_{2}}_{a_{3}a_{4}},\\
\Gamma^{(t)a_{1}a_{2}}_{a_{3}a_{4}}(k,k',\omega)=&\left[\frac{-1}{2}{\bf U}^{(s)}\chi^{(s)}(k-k',\omega)\chi^{(s)}-\frac{1}{2}{\bf U}^{(c)}\chi^{(c)}(k-k',\omega)\chi^{(c)}
+\frac{1}{2}{\bf U}^{(s)}+\frac{1}{2}{\bf U}^{(c)}\right]^{a_{1}a_{2}}_{a_{3}a_{4}},
\end{aligned}
\end{equation}
respectively.

The real static polarization in this case is independent of the frequency and it's proportional to the inverse bare Coulomb coupling as $1/g_{0}$ when
in the absence of static dielectric function, i.e., $\epsilon^{-1}=\epsilon_{0}(1+g({\bf k})\Pi({\bf k},0))=0$.
The dielectric function here is contributed by the electron-electron interaction (within RPA) 
$V=\Pi({\bf k},\omega)/\epsilon^{-1}({\bf k},\omega)$.
The high harmonic radiation intensity is
\begin{equation} 
\begin{aligned}
I(\omega)=|i\omega J(\Omega)|^{2},
\end{aligned}
\end{equation}
where $J(\omega)$ here is the Fourier transformation of the intraband current $J(\tau)=n_{{\bf k}}v_{\alpha(\beta)}$ for the electron channel or the hole channel,
through $J(\omega)=\int^{\pi/a}_{-\pi/a}J(t)e^{i\Omega t}$,
and the frequency-dependent interband polarization is absent here.
The intraband current has $J(\omega)=\frac{1}{\hbar}\int^{\pi/a}_{-\pi/a}n_{{\bf k}}(\omega)\frac{\partial}{\partial {\bf k}}\varepsilon({\bf k})$.

The real part and imagniary part of the free polarization function (U=0) can be related by the Kramers-Kronig relation\cite{Lucarini V}:
\begin{equation} 
\begin{aligned}
{\rm Re}[\Pi({\bf q},\Omega)]=\frac{2}{\pi}\mathcal{P}\int^{\infty}_{0}d\omega\frac{\omega{\rm Im}[\Pi({\bf k},\omega)]}{\omega^{2}-\Omega^{2}},\\
{\rm Im}[\Pi({\bf q},\Omega)]=-\frac{2\Omega}{\pi}\mathcal{P}\int^{\infty}_{0}d\omega\frac{{\rm Re}[\Pi({\bf k},\omega)]}{\omega^{2}-\Omega^{2}}.
\end{aligned}
\end{equation}

\section{Scattering due to charged impurty}

The impurities scattering potential after the Fourier transformation is $V({\bf k}_{s})=\frac{2\pi e^{2}}
{\epsilon_{0}\epsilon\sqrt{({\bf q})^{2}+{\bf k}_{s}^{2}}}$
with the screening wave vector ${\bf k}_{s}=2\pi e^{2}\Pi({\bf q},\omega)/(\epsilon_{0}\epsilon)$ which is polarization-dependent.
The effective Coulomb interaction with the effect of impurity is 
\begin{equation} 
\begin{aligned}
g_{{\rm eff}}({\bf q})=\frac{\frac{2\pi e^{2}}{\epsilon_{0}\epsilon\sqrt{{\bf }^{2}+{\bf k}_{s}}}}{1+\frac{2\pi e^{2}\Pi({\bf q},\omega)}{\epsilon_{0}\epsilon}},
\end{aligned}
\end{equation}
which can also be written as\cite{Ando T2}
\begin{equation} 
\begin{aligned}
g_{{\rm eff}}({\bf q})=\frac{\frac{2\pi e^{2}{\rm e}^{-{\bf q}r}}{\epsilon_{0}\epsilon{\bf q}}}{1+\frac{2\pi e^{2}\Pi({\bf q},\omega)}{\epsilon_{0}\epsilon}}
\end{aligned}
\end{equation}
The ${\bf q}$ is zero only in the case of elastic backscattering in the low-temperature limit 
where the scattering potential is close to a $ m_{D}$-function
similar to the Lorentzian representation
and become ${\bf q}$-independent.
In this case, the scattering potential is decay as $1/|{\bf k}_{s}|$.
Due to the exist of the impurities and lattice defects, the quantum spin-Hall effect with the spin-polarized current 
may more observable due to the SOC with the impurities
and it's robust against the nonmagnetic impurity scattering.

Taking into consider the Coulomb scattering by the charged impurity and with the non-static dielectric function,
the polarization function in one loop approximation (electron-hole bubble diagram) which 
containing both the intreband part and the intraband part can be wriiten as\cite{Hwang E H,Sensarma R,Ando T,Wunsch B,Tabert C J2}
\begin{equation} 
\begin{aligned}
\Pi({\bf q},\Omega)=-g_{s}g_{v}\frac{2\pi e^{2}}{\epsilon_{0}\epsilon}\sum_{m_{D}}
\int_{1st BZ}\frac{d^{2}k}{(2\pi)^{2}}\sum_{{\bf q};s,s'=\pm 1}\frac{f_{s({\bf k}+{\bf q})}-f_{s'{\bf k}}}{sE_{{\bf k}+{\bf q}}
-s'E_{{\bf k}}-\Omega-i\delta}{\bf F}_{ss'}({\bf k},({\bf k}+{\bf q})),
\end{aligned}
\end{equation}
where the factor $g_{s}g_{v}=4$ in the numerator denotes the spin and valley degenerates (or degrees of freedom),
$s,s'$ are the band index ($ss'=1$ for the intraband case and $ss'=-1$ for the interband case),
and the index of Dirac mass $m_{D}$ indicates the summation over the valley, spin, and pseudo spin degrees of freedom.
$f$ is the Fermi-Dirac function which can be estimated as step function in the zero temperature limit,
i.e., $f=1$ for the filled valence band and the flat band part 
which is possible for the silicene in spin-polarized semimetal phase\cite{Wu C H} and $f=\Theta({\bf k}_{F}-{\bf k})$ for the conduction band.
$E_{{\bf k}}$ is the energy (eigenvalue) of electron state and the spatial dependence is neglected here.
The energy of electron states is $\sim\hbar v_{F}{\bf k}$ when local around the Dirac-point
and $\sim h^{2}{\bf k}^{2}$ for the parabolic spectrum like for the AA-stacked bilayer silicene\cite{Wu C H}.
The transported momentum ${\bf q}$ is zero only for the elastic backscattering in which case the scattering potential is close to a $ m_{D}$-function
similar to the Lorentzian representation
and become $ m_{D} {\bf k}$- and ${\bf k}_{s}$-independent,
in which case the scattering potential is decay as $1/|{\bf k}_{s}|$ with ${\bf k}_{s}$ the scattering wave vector.
Distinct from Refs.\cite{Ando T,Wunsch B,Sensarma R,Kotov V N},the Coulomb interaction matrix element 
\begin{equation} 
\begin{aligned}
{\bf F}_{ss'}({\bf k},({\bf k}+{\bf q}))=ss'{\rm cos}^{2}\theta_{\sigma\eta}=\frac{1}{2}\left[1+ss'(\frac{{\bf k}({\bf k}+{\bf q})}{E_{{\bf k}}E_{{\bf k}+{\bf q}}}
+\frac{4m_{D}^{2}}{E_{{\bf k}}E_{{\bf k}+{\bf q}}})\right],
\end{aligned}
\end{equation}
where the angle $\theta_{\sigma\eta}={\rm arctan}\frac{\eta\hbar v_{F}{\bf k}}{2m_{D}}$ is defined in the scattering phase space (see Fig.1(d)-(e))
where the Dirac-mass is taken into consider.
In Fig.1(e), the intervalley scattering which must be anisotropic 
is possible through the edge states.
Here the Dirac-mass is related to the band gap in Dirac-cone by $2m_{D}= \Delta$\cite{Wu C H2}.
While in the simply circular coordinate system,
the angle $\theta$ between ${\bf k}$ and ${\bf k}+{\bf q}$ has
${\rm cos}\theta=\langle\chi({\bf k})|\chi({\bf k}+{\bf q})\rangle=(k+q{\rm cos}\phi)/\sqrt{k^{2}+q^{2}+2kq{\rm cos}\phi}$ 
where $\phi$ is the angle between ${\bf k}$ and ${\bf q}$, and $|\chi({\bf k})\rangle=\psi_{s}^{*}({\bf k})\psi_{s'}({\bf k})$,
$|\chi({\bf k}+{\bf q})\rangle=\psi_{s}({\bf k}+{\bf q})\psi_{s'}^{*}({\bf k}+{\bf q})$ is the eigenstates with the eigenvectors $\psi$ of the Hamiltonian.
Note that here the scalar product are the simplification of $\langle\chi({\bf k})|\int^{\pi/a}_{-\pi/a}e^{-i{\bf q}r{\rm cos}\theta}d\theta|{\bf k}'\rangle=
\langle\chi({\bf k})|\chi({\bf k}')\rangle \delta({\bf k}',{\bf k}+{\bf q})$.
The scattering angle has ${\bf q}=|{\bf k}-{\bf k}'|=2k\ {\rm sin}\theta$\cite{Adam S,Vargiamidis V}, where $\theta$
describes the difference between the monentums before scattering and after scattering,
and it tends to zero ($\theta\rightarrow 0$) only in the case of SC silicene 
(deposited on a SC electrode or generate the topological superconductor by the STM probe).

The polariztion function will becomes $\omega$-independent for the interband transition\cite{Hwang E H} between the conduction band and valence band
which only happen in the strong Coulomb-coupling case in the monolayer silicene\cite{Wu C H}.
Here we comment that, the analytical continuation here won't make the relations\cite{Pyatkovskiy P K}: $\Pi({\bf q},-\omega)=\Pi^{*}({\bf q},\omega),
{\rm Im}(1/i\omega)=-\pi m_{D}(\omega)$
lose efficacy, which are always valid in the nonstatic case ($\omega> 0$).

Here the scattering term of the polarization function is distinct from that of the monolayer MoS$_{2}$\cite{Scholz A},
which connects the two states (before and after scattering) by the scalar product of eigenstates.
Within the process of transition (described by the scalar product) of monolayer MoS$_{2}$,
the spin and valley index won't change and only the change of pseudospin is possible,
while for silicene,
the interband transitions through the edge states (like the helical edge state which with the up- and down-spin flow toward the opposite directions
and the chiral edge which with the up- and down-spin flow toward the same direction in each edge)
provide the possibility for the spin-flip and valley-change (K to K') during the interband scattering.

For $\mu< m_{D}$, the polarization function in QED$_{2+1}$ can be written as\cite{Appelquist T W}
\begin{equation} 
\begin{aligned}
\Pi({\bf q},\omega)=-g_{s}g_{v}\frac{e^{2}\hbar^{2}v_{F}^{2}{\bf q}^{2}}{2\epsilon_{0}\epsilon(\hbar^{2}v_{F}^{2}{\bf q}^{2}-\hbar^{2}\omega^{2})}
(2m_{D}+\frac{\hbar^{2}v_{F}^{2}{\bf q}^{2}-\hbar^{2}\omega^{2}-4m_{D}^{2}}{\sqrt{\hbar^{2}v_{F}^{2}{\bf q}^{2}-\hbar^{2}\omega^{2}}}{\rm arcsin}
\sqrt{\frac{\hbar^{2}v_{F}^{2}{\bf q}^{2}-\hbar^{2}\omega^{2}}{\hbar^{2}v_{F}^{2}{\bf q}^{2}-\hbar^{2}\omega^{2}+4m_{D}^{2}}}),
\end{aligned}
\end{equation}
In QED$_{2+1}$, the mass has $m_{D}\sim (T/c)^{2}\sqrt{\hbar v_{F}{\bf q}}$\cite{Sachdev S} in the long-wavelength case, where $c$ is the speed of light
and $v_{F}=c$ here.
It's found that the QED$_{2+1}$ theory is power tool for the undopped silicene, graphene and other two-dimension materials,
but for the case of finite chemical potential, i.e., for the dopped case,
especially for the case of chemical potential larger than the Dirac-mass,
the resulting broken of the electron-hole symmetry since it's far away from the half-filling,
may makes it lose efficacy,
except under the Feynman gauge with Lorentz-invariance, where the current has $\partial_{\mu}{\bf J}^{\mu}=0$
and thus satisfy the continuity equation $\frac{\partial\rho}{\partial t}+\nabla\cdot {\bf J}$ where $\rho$ is the probability density.
The electromagnetic potential $A$ also need to obeys the $\partial_{\mu}{\bf A}^{\mu}=0$, and
the electromagnetic coupling-related causal retarded propagator (especially in the nonrelativistic limit which with $c\rightarrow\infty$ and $m_{D}\rightarrow 0$
($m^{*}\rightarrow 0$))
with lorentz invariance by the time ordered product is
\begin{equation} 
\begin{aligned}
G(t'-t)=-i\langle\mathcal{T}{\bf A}^{\mu}(t){\bf A}^{\mu}(t')\rangle\\
=\int\frac{d^{2}q}{(2\pi)^{2}}\frac{d\omega}{2\pi}\frac{e^{-i\omega(t'-t)]}}{\omega^{2}-\varepsilon^{2}+i\eta_{s}}.
\end{aligned}
\end{equation}

Base on the above polarization function with the scattering by the charge or spin fluctuations and
in the case of small chemical potential $\mu< m_{D}$,
the purely real polarization is achieved if $\omega<\sqrt{{\bf q}^{2}+4 m_{D}^{2}}$,
which is static polarization now due to the absence of imaginary part of polarization function.
In the case of large chemical potential (thus large density) 
$\mu> m_{D}$, the polarization function becomes frequency-dependent,
as shown in the Fig.4(a)-(b), the region 1A and 2B are correspond to $\hbar\omega<\mu-\sqrt{\hbar^{2}v_{F}^{2}({\bf q}-{\bf k}_{F})^{2}+ m_{D}^{2}}$
and $\mu+\sqrt{\hbar^{2}v_{F}^{2}({\bf q}-{\bf k}_{F})^{2}+ m_{D}^{2}}<\hbar\omega<\mu+\sqrt{\hbar^{2}v_{F}^{2}({\bf q}+{\bf k}_{F})^{2}+ m_{D}^{2}}$,
respectively (see Refs.\cite{Tabert C J2,Pyatkovskiy P K,Wunsch B,Ando T,Tabert C J2,Pyatkovskiy P}).
Specially, in A region, the polarization function only has the imaginary part, but the the imaginary polarization is also vanish in the point $\omega=0$
(static polarization) as shown in the Fig.4(b)-(c),
thus the purely real polarization can be achieved when $\omega=0$ in the 1A regions or $\omega=\omega_{c}$ in the 2B regions,
where the such critical value $\omega_{c}$ is shown in the Fig.4 for the case of $\mu=1$ and $\mu=2$.
It also reveal that, the imaginary part of polarization function may not vanishes even it's static ($\omega=0$)
when $\mu> m_{D}$.
From Fig.4(b)-(c),
we can see that the static polarization is vanishes in 1A region, while in the 3A region where 
the scattering is exceed to the Fermi surface and 
in the region, 
$\hbar\omega<-\mu+\sqrt{({\bf q}+{\bf k}_{F})^{2}+ m_{D}^{2}}$ (i.e., ${\bf q}>2{\bf k}_{F}$ for the static case).
The singular point for the static dielectric function are indicated as 2${\bf k}_{F}$ in Fig.4(c)-(d) where the static polarization function
shows discontinuity in the first derivative,
and corresponds to the points with ${\bf k}={\bf k}+{\bf q}={\bf k}_{F}$ (like the elastic backscattering) with $\langle {\bf k},{\bf q}\rangle=\pi$\cite{Lin M F}.
The polariztion function is also isotropic within the range of ${\bf q}\le 2{\bf k}_{F}$\cite{Liu Y}.
We obtain the same results with the previous literatures\cite{Chang H R,Pyatkovskiy P K}: 
for static case ($\omega=0$), the polarization function diverges ar ${\bf q}=2{\bf k}_{F}$,
i.e., the it's first derivative is discontinuous at these points,
and for gapless case ($ m_{D}=0$), the first derivative is continuous but the discontinuity appears for the second derivative.

For this case, the static polarization must be a real quantity and it's continuous through all the long-wavelength regime
(see Fig.3).
The expressions of static polarization at zero temperature are given as\cite{Gorbar E V,Pyatkovskiy P K}
\begin{equation} 
\begin{aligned}
\Pi({\bf q},0)=-g_{s}g_{v}\frac{2e^{2}\mu}{2\pi\epsilon_{0}\epsilon \hbar^{2} v_{F}^{2}}
\left[\frac{ m_{D}}{2\mu}+\frac{{\bf q}^{2}-4 m_{D}^{2}}{4\hbar v_{F}{\bf q}\mu}{\rm arcsin}\sqrt{\frac{\hbar^{2}v_{F}^{2}{\bf q}^{2}}{\hbar^{2}v_{F}^{2}{\bf q}^{2}+4 m_{D}^{2}}}\right]
\end{aligned}
\end{equation}
for $0<\mu< m_{D}$,
\begin{equation} 
\begin{aligned}
\Pi({\bf q},0)=-g_{s}g_{v}\frac{2e^{2}\mu}{2\pi\epsilon_{0}\epsilon \hbar^{2} v_{F}^{2}}
\left[1-\Theta({\bf q}-2{\bf k}_{F})\left(\frac{\sqrt{{\bf q}^{2}-4{\bf k}_{F}^{2}}}{2{\bf q}}-\frac{\hbar^{2}v_{F}^{2}{\bf q}^{2}-4 m_{D}^{2}}{4\mu \hbar v_{F}{\bf q}}{\rm arctan}\frac{\hbar v_{F}\sqrt{{\bf q}^{2}-4{\bf k}^{2}_{F}}}{2\mu}\right)\right]
\end{aligned}
\end{equation}
for $\mu> m_{D}$.
The polts of static polarization is shown in the Fig.3.

The regions with different characteristics of the polarization and with finite frequency can be specified by the relations between the frequency and the
scattering momentum, e.g.,
the regions below $\omega=2\mu$ are defined by\cite{Pyatkovskiy P K,Tabert C J2} (see Fig.9):
\begin{equation} 
\begin{aligned}
&1B:\ \hbar v_{F}{\bf q}<2{\bf k}_{F},\ \sqrt{\hbar^{2}v_{F}^{2}{\bf q}^{2}+4 m_{D}^{2}}<\hbar\omega<\mu+\sqrt{\hbar^{2}v_{F}^{2}({\bf q}-{\bf k}_{F})^{2}+ m_{D}^{2}},\\
&5B:\ \hbar v_{F}{\bf q}<\hbar\omega<\sqrt{\hbar^{2}v_{F}^{2}{\bf q}^{2}+4 m_{D}^{2}},\\
&4A:\ -\mu+\sqrt{\hbar^{2}v_{F}^{2}({\bf q}^{2}+{\bf k}^{2}_{F})+ m_{D}^{2}}<\hbar\omega<\hbar v_{F}{\bf q},\\
&2A:\ \pm\mu\mp\sqrt{\hbar^{2}v_{F}^{2}({\bf q}-{\bf k}_{F})^{2}+ m_{D}^{2}}<\hbar\omega<-\mu+\sqrt{\hbar^{2}v_{F}^{2}({\bf q}^{2}+{\bf k}^{2}_{F})+ m_{D}^{2}},\\
&3A:\ \hbar\omega<-\mu+\sqrt{({\bf q}+{\bf k}_{F})^{2}+ m_{D}^{2}},\\
&2B:\ \mu+\sqrt{\hbar^{2}v_{F}^{2}({\bf q}-{\bf k}_{F})^{2}+ m_{D}^{2}}<\hbar\omega<\mu+\sqrt{\hbar^{2}v_{F}^{2}({\bf q}^{2}+{\bf k}^{2}_{F})+ m_{D}^{2}},\\
\end{aligned}
\end{equation}
with the Fermi wave vector ${\bf k}_{F}=\sqrt{\mu^{2}- m_{D}^{2}}$ in scattering phase space (which consider the degrees of freedom $\sigma$ and $\eta$;
see Fig.1(d)-(e)).
The polarization function is presented in Fig.4 where the factors $\hbar$ and $v_{F}$ are setted as 1 for simplicity.
Distincted from the static results presented in Fig.4(c)-(d) where the discontinuity is only emerge in ${\bf q}=2{\bf k}_{F}$ (here we only take into account
the Dirac gap formed by the up-spin bands, i.e., there are only one gap and Fermi wave vector in each cone),
the discontinuity of nonstatic polarization is arounds the ${\bf q}=\omega$ (see Fig.4(e)-(j)).
From Fig.4, we found that the polarization is continuous although has a abrupt peak in the discontinuous point of the first derivative,
and the static polarization is purely real.
The effects from the Rashba-couping and the exchange-field for the silicene can also be taken into accout by the Dirac-gap
$m_{D}=\eta\lambda_{{\rm SOC}}s_{z}-\frac{\overline{\Delta}}{2}E_{\perp}+Ms_{z}$,
as done in Ref.\cite{Scholz A2} for graphene.

For the large gap $ m_{D}=1.148$ eV in Fig.4(e), discontinuous point for the first derivative is at ${\bf q}<\sqrt{16- m_{D}^{2}}-{\bf k}_{F}=2.194$
which is labeled in the figure.
By comparing Fig.4(e) and (f), we can obtain that the region 4$A$ is decrease with the decreasing gap, and will vanishes for the gapless case.

In Fig.4 we only consider the effects of the electric field and the intrinsic SOC which are the mainly effects on silicene,
but there are also some other effects which can slightly affects the band gap,
like the NN (induced by electric field) and NNN Rashba-coupling or the exchange field (including the spin-dependent part and the
charge-dependent part), or even the light (electromagnetic wave) in a certain frequency\cite{Wu C H2}.
Among these effect, we have found that the electric field-induced NN Rashba-coupling is proportional to the applied perpendicular electric field (see Ref.\cite{Wu C H})
as $R_{2}(E_{\perp})=0.012E_{\perp}$ where the electric field is in unit of meV/\AA\ here.
In Fig.5, we  make a comparation for the band structures near Dirac-cone with and without consider the effect of Rahsba-coupling and exchange field.
From Fig.5(b), we find that, consider the electric-field-induced Rashba-coupling, the up-spin bands behave like the down-spin bands in Fig.5(a),
and the band gap is closed until approaches the critical value $E_{\perp}=E_{\perp c}=0.017$ eV.
For $E_{\perp}>E_{\perp c}$, the evolution of band gap is the same as the (a) case.
Taking both the Rashba-coupling and the exchange field into consider, the symmetry between conduction band and valence band reappear (see Fig.5(c)).

The Dirac point approximation which ignores the trigonal warping term due to the anisotropic trigonal Fermi energy contours 
has been explained in one of our other works (see Fig.4 of Ref.\cite{Wu C H}), and thus it's only valid for the small ${\bf q}$ case (long-wavelength limit)
unlike the single Dirac cone approximation\cite{Farajollahpour T},
In Dirac point approximation,
the static screened Coulomb potential of the charged impurity can be obtained by RPA as\cite{Tabert C J2,Scholz A,Chang H R,Malcolm J D}
\begin{equation} 
\begin{aligned}
\Phi(r)=&\frac{2\pi Q}{\epsilon_{0}\epsilon}\int\frac{d^{2}q}{(2\pi)^{2}}\frac{e^{-i{\bf q}r}}{\epsilon({\bf q},0){\bf q}}\\
=&\frac{Q}{\epsilon_{0}\epsilon}\int^{\infty}_{0}d{\bf q}\frac{J_{0}({\bf q}r)}{\epsilon({\bf q},0)},
\end{aligned}
\end{equation}
where $Q$ is the charge of impurity, $J_{0}({\bf q}r)$ is the zeroth Bessel function of the first kind, 
the zeroth index here is due to the two-dimensional Lindhard function\cite{Béal-Monod M T},
$\epsilon({\bf q},0)$ is the static dielectric function. The screened charge density is\cite{Wunsch B} 
$n(r)=\frac{Q}{4\pi^{2}}\int d{\bf q}e^{i{\bf q}\cdot{\bf r}}(\epsilon^{-1}({\bf q},0)-1)$.
Through Fourier transform, we can obtain $\Phi({\bf q})=g({\bf q})Q/\epsilon({\bf q},0)$.
Deffer from the screened potential, the screened spin or charge density by the Coulomb repulsion is decay as $r^{-3}$
in large distance, rather that $r^{-2}$.
As a example, the $r^{-3}$-decay was found in the density of states of the graphene when away from the van Hove singularities\cite{Vozmediano M A H},
which are the points with largest density of states and corresponds to the $M$-point of the Brillouin zone.
However, the induced charge density is inversely proportional to the chemical potential since its fluctuation is dominated by the plasmon model,
while the spin density is not.
The static dielectric functions in the case of long wavelength and short wavelength are
\begin{equation} 
\begin{aligned}
\epsilon({\bf q},0)=\left\{
\begin{array}{rcl}
1+2\pi e^{2}\Pi({\bf q},\omega)/(\epsilon_{0}\epsilon {\bf q}),&\ {\rm for}\ \hbar v_{F}{\bf q}< 2{\bf k}_{F},\\
1+\frac{g_{s}g_{v}\pi }{8}r_{w},&\ {\rm for}\ \hbar v_{F}{\bf q}> 2{\bf k}_{F},
\end{array}
\right.
\end{aligned}
\end{equation}
where the Wigner-Seitz radius $r_{w}$
is a dimensionless constant for the sattering potential-independent case,
but becomes impurity concentration- and electron density (band filling)-dependent when with the charged impurity or the electron liquid, respectively
\cite{Hwang E H2},
and it can be controlled by turning the gate voltage.
For the varied Fermi wave vector in the above static case, the static polarization-dependent Wigner-Seitz radius has a form distinct from the effective fine structure constant
which is $e^{2}/\epsilon_{0}\epsilon\hbar v_{F}$
\cite{Tabert C J2,Malcolm J D}: $r_{w}
=\frac{e^{2}\pi}{2\epsilon_{0}\epsilon {\bf k}_{F}}\Pi({\bf q},0)=\frac{e^{2}}{\epsilon_{0}\epsilon\hbar\gamma}$
where the band parameter $\gamma=2{\bf k}_{F}/(\pi\Pi({\bf q},0)\hbar)$($\sim v_{F}$ for the monolayer silicene with large carriers density) 
is inversely proportional to the static polarization function.
The short-wavelength ($\hbar v_{F}{\bf q}> 2{\bf k}_{F}$) 
behavior also rised with the enhanced of the interband transition or polarizability, and thus related to the longitudunal conductivity.

For large $r$ case, the screended potential as well as the induced charge or spin density mainly
contain two parts of the contricution.
The first part is Thomas-Fermi contribution in long-wavelength approximation (consistent with the RPA for ${\bf q}\rightarrow 0$)
with large Wigner-Seitz radius $r_{w}$.
The Thomas-Fermi decay of $\Phi(r)$ scale as $1/r^{3}$ when it with nonzero Dirac-quasiparticle scattering rate and nonzero temperature.
It's also show that the Thomas-Fermi wave vector could not depends on the momentum (including ${\bf k}_{F}$) or frequency 
and thus with the static polarization in the absence of Dirac-quasiparticle scattering but with nonzero temerature\cite{Gorbar E V}:
\begin{equation} 
\begin{aligned}
\Pi(0,0)=-g_{s}g_{v}\frac{e^{2}T}{\pi\epsilon_{0}\epsilon v_{F}^{2}}\left[{\rm ln}(2{\rm cosh}\frac{m_{D}+\mu}{T})-\frac{m_{D}}{2T}{\rm tanh}\frac{m_{D}+\mu}{2T}+(\mu\rightarrow -\mu)\right],
\end{aligned}
\end{equation}
then for the zero-temperature case, it becomes proportional to the density of states $D$ of the Dirac-quasiparticle which is a step function now:
\begin{equation} 
\begin{aligned}
\Pi(0,0)_{T\rightarrow 0}=
-e^{2}D(|\mu|)=-e^{2}\frac{g_{s}g_{v}|\mu|}{2\pi\hbar^{2}v_{F}^{2}}\frac{1}{2}\sum_{\eta=\pm 1}\left[\theta(|2\mu|-2|m_{D}|_{\eta})\right].
\end{aligned}
\end{equation}
The second part is the Friedel oscillation\cite{Friedel J} in the next order which is anisotropic due to the anisotropic dielectric function\cite{Farajollahpour T}
and it's only exist in the case of $\mu> m_{D}^{{\rm max}}$,
i.e., when $\mu< m_{D}^{{\rm min}}$ or $ m_{D}^{{\rm max}}>\mu> m_{D}^{{\rm min}}$,
the screened potential won't shows the oscillation behavior.
It's also found that the oscillation of the screended potential vanishes for the large chemical potential\cite{Pyatkovskiy P K2},
just like the light away of the beating for the Friedel oscillatory in large $\mu$ case (see Fig.6(a)-(b)).

The Friedel oscillatory decay of screended potnetial is scale as ${\rm sin}(2{\bf k}_{F}r)/r^{2}$ for the gapped silicene or graphene, 
and the relativistic two-dimension electron gas (2DEG)
\cite{Stern F,Pyatkovskiy P K}
especially for the large distance $r$,
while for short distance,
it decays as ${\rm cos}(2{\bf k}_{F}r)/r^{3}$ like the gapless silicene or graphene which with ${\bf k}_{F}^{{\rm max}}=\mu$
due to the constant nontrivial Berry phase ($\pi$ for the monolayer one and $2\pi$ for the bilayer one\cite{Park C H}),
and the traditional nonrelativistic two-dimension electron gas (2DEG).
Here we comment that, 
in long-wavelength limit, even the gapped silicene shows the oscillatory decay scale as $r^{-3}$ (as shown in the Fig.6(c),(f),(i)).

The power-law-dependent Friedel oscillation can be enhanced by increasing the Rashba-coupling (through the impurity adatoms' surface
deposition or by increasing the on-site Hubbard U) even for the graphene\cite{Scholz A2}.
That's due to the increasing of Rashba-coupling can enlarge the difference of curvature between the conduction band and valence band (see Fig.5(b)),
and hence enhance the Friedel oscillation.
The results about the Friedel oscillation contribution are presented in Fig.6,
for the large distance behavior.
For large chemical potential $\mu=2$, as shown in Fig.6(a)-(b), the beating phenomenon is not obvious for the large distance decay of screened potential,
but it's obviously for the smaller chemical potential (see other pannels).
The relaxation of $\Phi(r)$ is towards the zero no matter what value the chemical potential and electric field is.
As shown in the Fig.6(g)-(l), when without electric field ($E_{\perp}=0$ eV), 
the spin degenerate as well as the symmetry between the lowest conduction band and highest valence band is keeped, 
in this case, $ m_{D}^{{\rm max}}= m_{D}^{{\rm min}}=0.0078$ eV and thus ${\bf k}_{F}^{{\rm min}}={\bf k}_{F}^{{\rm max}}=1.2$ where we set $\mu=2$ here,
then the Friedel oscillation becomes single-component and the beating of Friedel oscillation is vanishes (see Fig.6(k)).
The beating also vanishes 
when $E_{\perp}=0.017$ eV (i.e., $ m_{D}^{{\rm min}}=0$ which correpons to the ${\bf k}_{F}^{{\rm max}}=\sqrt{\mu^{2}-\lambda_{SOC}^{2}}$)
in which case that the screened potential is consisted of two types of decaying: ${\rm cos}(2{\bf k}_{F}r)/r^{3}$ and ${\rm sin}(2{\bf k}_{F}r)/r^{2}$,
and the $r^{-2}$ one is dominate for the large distance case,
and thus exhibits non-beating behavior (as shown in the Fig.6(m)-(n)) just like the case of $E_{\perp}=0$.
That also indicates the phase transition point from the gapless semimetal to the gapped band insulator.

\section{plasmon branch in collective model}
The above-mentioned long-wavelength behavior with small ${\bf q}$ (${\bf q}\ll\omega\ll\mu$) also related to the plasma physics 
(a collective model of the oscillating electrons)
due to its unstable nature.
In a more macroscopic view, the above occupation satisfy $n_{h}=n_{e}+n_{p}$ in the charge neutrality case
where the $n_{p}$ is the occupation of the plasmon in a form of dust grain\cite{Lampe M},
and here the oscillation frequency of the dusty plasmon is $\omega_{p}=\sqrt{\frac{e^{2}n_{p}}{\epsilon_{0}\epsilon m_{d}}}$ 
where $m_{p}$ is the mass of the dusty plasmons. 
For the collective model of the electron-ion plasma\cite{Saleem H,Wunsch B}
which disturbed by the disorder from the stream instability,
the resulting plasmon polarization is $\Pi({\bf q},\omega_{p})=\Pi({\bf q},\omega_{ele})+\Pi({\bf q},\omega_{ion})$,
where $\omega_{ion}=\sqrt{\frac{e^{2}n_{ion}}{\epsilon_{0}\epsilon m_{ion}}}$ 
is the oscillation frequency of ions with the ion mass $m_{ion}$ and the $\omega_{ele}$ is frequency of electron.
Note that the oscillation of ion here is possible in the case of zero screening of the long-range Coulomb interaction 
by the charged impurity or the conduction electrons as we discussed in Ref.\cite{Wu C H},
since the screening effect will suppress the opening of the band gap and breaks the Coulomb long-range order.
As mentioned in one of our early works\cite{Wu C H2}, where
we have applied the off-resonant circular polarized laser beam on the silicene sample,
which with the frequency $\omega\gg t$ and can be relativistic self-focusing, thus overcome its diffraction when it's in the plasma channel
with the plasma frequency is $\omega_{p}=\frac{2eE_{F}}{\hbar}\sqrt{\frac{1}{3\pi\hbar v_{F}\epsilon_{0}\epsilon_{s}}}$,
thus it's off-resonant where the electrons cannot directly
absorb the photons\cite{Kitagawa T},
and the electron effective mass can be controlled by the laser intensity by drive the electrons to quiver with a determined velocity,
which is similar to the case of vertical electric field.

In our tight-binding model, the plasmon dispersion which is related to the Fermi energy ($\sim\sqrt{\frac{e^{2}{\bf q}E_{F}}{\epsilon_{0}\epsilon}}$) 
in the long-wavelength limit, 
can be determined the zeros of the dielectric function 
(or the pole of the energy loss function as shown in Fig.7(b))\cite{Wunsch B}: $\epsilon({\bf q},\omega_{p}-i\nu)=0$.
The plasmon decay rate 
$\nu=\frac{{\rm Im}[\Pi({\bf q},\omega_{p})]}{\frac{\partial}{\partial \omega}{\rm Re}[\Pi({\bf q},\omega_{p})]}$ 
is nonzero in the single-particle excitation (electron-hole continuum) regime 
where ${\rm Im}[\Pi({\bf q},\omega_{p})]\neq 0$\cite{Chang H R}.
Thus the lifetime of the damped plasmon $\tau^{-1}$ is proportional to the imaginary part of the polarization function 
as $\tau^{-1}\propto -{\rm Im}\Pi({\bf q},\omega_{p})$. 
The acoustic phonons decay rate also rised in this regime by the electron-hole excitation.
The long-wavelength plasmon frequency 
of the monolayer silicene in nonlocal case is
\begin{equation} 
\begin{aligned}
\omega_{p}=\sqrt{g_{s}g_{v}}\sqrt{\frac{e^{2}{\bf q}\mu}{2\epsilon_{0}\epsilon\hbar v_{F}}\left[2-\frac{( m_{D}^{{\rm max}})^{2}+( m_{D}^{{\rm min}})^{2}}{\mu^{2}}\right]},\ {\rm for}\ 
{\bf q}> m_{D}^{{\rm max}},\\
\omega_{p}=\sqrt{g_{s}g_{v}}\sqrt{\frac{e^{2}{\bf q}\mu}{\epsilon_{0}\epsilon\hbar v_{F}}\left[1-\frac{( m_{D}^{{\rm min}})^{2}}{\mu^{2}}\right]},\ {\rm for}\ 
 m_{D}^{{\rm max}}>{\bf q}> m_{D}^{{\rm min}},
\end{aligned}
\end{equation}
which is in a nonclassical form and $\sim 1/\sqrt{\hbar}$ like the monolayer graphene\cite{Sarma S D}
In fact, for the long-wavelength plasmon dispersion, both the monolayer and bilayer silicene and the 2DEG, 
have a similar form which is $\sim \sqrt{\frac{g_{s}g_{v}e^{2}{\bf q}\mu}{\epsilon_{0}\epsilon\hbar v_{F}}}$\cite{Sensarma R,Stern F,Kotov V N}.
Note that this universal relation requires the low-temperature condition ($T<\mu$).
For the nonlocal long-wavelength case, the nonclassical plasmon dispersion of bilayer silicene with quantum correlations reads\cite{Sensarma R}
\begin{equation} 
\begin{aligned}
\omega_{p}=\sqrt{g_{s}g_{v}}\sqrt{\frac{e^{2}{\bf q}E_{F}}{\epsilon_{0}\epsilon\hbar v_{F}}\left(1-\frac{r_{w}{\bf q}}{8{\bf k}_{F}}\right)},
\end{aligned}
\end{equation}
which is Wigner-Seitz radius $r_{w}$-dependent and has
$\omega_{p}\sim n^{1/2}$ since the Fermi energy $E_{F}=\hbar^{2}{\bf k}_{F}^{2}/2m^{*}\sim n$
with the effective electron mass $m^{*}$ about the interlayer motion,
and $r_{w}\sim n^{-1/2}$ here similar to the two-dimension electron liquid\cite{Hwang E H2} or 2DEG.
It's deffer from the plasmon dispersion of the monolayer silicene or graphene or MoS$_{2}$ or WS$_{2}$
\cite{Gamayun O V,Falkovsky L A} which is $\omega_{p}\sim n^{1/4}$
since the $\gamma$ is treated as a constant now and the Fermi energy is $E_{F}=\gamma{\bf k}_{F}$.
The electron density-dependent Fermi wavevector has ${\bf k}_{F}=\sqrt{\pi n}\sim n^{1/2}$ with the filling density (carriers) $n=E_{F}/(\pi \hbar^{2}v_{F}^{2})$
for the two-dimwnsion system, 
and hence the Thomas-Fermi wavevector also has
${\bf q}_{{\rm TF}}=g_{s}g_{v}r_{w}{\bf k}_{F}\sim n^{1/2}$.
Since this expression of the bilayer silicene plasmon frequency is Fermi energy dependent,
the plasmon dispersion vanishes for the undoped (without the band filling and thus with zero density of states) intrinsic bilayer silicene 
which with zero Fermi energy and electron density as discussed in Ref.\cite{Hwang E H}.
For the bilayer silicene (or the multilayer bulk form like the graphene) within the local long-wavelength,
the classical plasmon dispersion is
\begin{equation} 
\begin{aligned}
\omega_{p}=\sqrt{g_{s}g_{v}}\sqrt{\frac{e^{2}{\bf q}\mu}{\epsilon_{0}\epsilon}e^{-ik_{z}d-{\bf q}d}},
\end{aligned}
\end{equation}
with the asymptotic factor $e^{-ik_{z}d-{\bf q}d}=\frac{{\rm sinh}({\bf q}d)}{{\rm cosh}({\bf q}d)-{\rm cos}(k_{z}d)}$\cite{Fetter A L}.
$k_{z}$ is the interlayer quasi-momentum which is within the range of $[-\pi/d,\pi/d]$ for the first Brillouin zone
and it's $k_{z}\ll 1/d$ in the long-wavelength limit, $d$ is the interlayer distance.
Then the classical plasmon dispersion can be obtained from the zeros of the bilayer dielectric function which is
$1+g({\bf q})\Pi({\bf q},\omega)e^{-ik_{z}d-{\bf q}d}$ where the Coulomb interaction $g({\bf q})=\frac{2\pi e^{2}}{\epsilon_{0}\epsilon{\bf q}}$.
Similarly, we can obatin the dielectric function and the potential function of the multilayer silicene,
e.g., the potential becomes $V({\bf q},\omega)=\delta_{ll'}+\sum_{l'}ge^{-{\bf q}|l-l'|d}\Pi({\bf q},\omega)V_{l'}({\bf q})$.
If consider the intraband transitions only in the low-frequency and low-temperature ($T<\mu$) regime,
the plasmon frequency is given as\cite{Falkovsky L A}
\begin{equation} 
\begin{aligned}
\omega_{p}=\sqrt{g_{s}g_{v}}\sqrt{\frac{2e^{2}T{\bf q}}{\hbar}{\rm ln}(2{\rm cosh}\frac{\mu}{2T})},
\end{aligned}
\end{equation}
which has been presented in Fig.8 with low-temperature $T=1$ K,
and it's obviously that it's well fits the classical bilayer silicene with $\mu=2$ in the low-frequency region.
That also consistent with the conclusion about the domination of the intraband transition in the low-frequency region.

The optical conductivity, energy loss function, and the dielectric function of the bilayer silicene are presented in the Fig.7.
The energy loss function ${\rm Im}[-1/\epsilon({\bf q},\omega)]$ provides the spectral density for the single-particle excitation regime,
and the damping in single-particle excitation regime also leads to the resonance of the energy loss process.
In fact, the process of the energy-loss is due to the intraband and intreband transition 
which also results in the losses of the density of states.

As shown in the Fig.8, the long-wavelength plasmon dispersion 
is obviously proportional to $\sqrt{{\bf q}}$ which consistent with the most two-dimension materials and 2DEG
and distincted from the high-energy $\pi$-plasmon model which is linear with ${\bf q}$.
We can see that the local long-wavelength plasmon dispersion is 
simply electric field (band gap)-dependent,
while for numerical result of RPA\cite{Shung K W K} which is valid even extend to the Wigner-Seitz radius $r_{w}$-dependent short-wavelength region
by solving the relation $\epsilon({\bf q},\omega_{p}-i\nu)=0$, which can be
further approximated as ${\rm Re}[\epsilon({\bf q},\omega_{p})]=0$
in the weak damping case (the result is presented in Fig.9)
with the plasmon dispersion much larger that $\nu$\cite{Scholz A},
thus the Landau-damping is ignored.
The $1B$ and $5B$ are outside the single-particle excitation regime
and thus polarization function in these regions is purely real (see Fig.4).

In the short-wavelength case,
the plasmon dispersion is not always simply $\sqrt{{\bf q}}$-dependent,
but changes suddenly (redshifted) once it enters the single-particle excitation regime\cite{Chang H R,Wunsch B}, 
like the interband single-particle excitation regime ($2B$)
or intraband single-particle excitation regime ($1A$ and $2A$) (see Fig.9),
where the imaginary part of the polarization function ${\rm Im}[\Pi({\bf q},\omega)]$ is nonzero
and thus the plasmon damped into the electron-hole pairs in these regions
due to the nonzero $\nu$,
and the acoustic phonons in 1$A$ region with long wavelength also exhibit such behavior\cite{Wunsch B}.
Through Fig.9 we can also see that the intraband transition is plays the leading role in the low frequency region 
with the acoustic plasmon model ($\sim {\bf q}$) and the interband transition
is plays the leading role in the high frequency region with the optical plasmon model ($\sim{\bf q}$).
In fact, for higher frequency (much larger than the threshold of interband transitions),
the optical propertices of silicene depends more on the fine structure constant that the frequency\cite{Falkovsky L A},
like the case of HHG.
The cause of damping in the large momentum region is mainly due to the intraband transition (${\bf q}>2{\bf k}_{F}$)
rather than the interband transition,
but for the dice lattices in massless Dirac model,
it's also reported\cite{Malcolm J D} that there exist a float-hebavior between the two cones in the 3$A$ region.
For the normal two-dimension materials or 2DEG, it obeys quadratic dispersion\cite{Hwang E H2} and
the acoustic branchs also exhibit such damped behaviors in the single-particle excitation regime.
A damped region between $2B$ and $1A,2A$ is possible by the interband transition between the two conduction bands in the case of finite $p$-doping
and taking into consider the effect of Rashba-coupling\cite{Scholz A2}.
This narrow region which haven't appeared in other literatures is critical for the 
relations between the intrinsic coupling and the Rashba-coupling,
and it need to obeys that the value of Fermi energy is larger than the plasmon energy, $|E_{F}|>\hbar\omega_{p}$\cite{Chen J},
otherwise the interband transition between two conduction bands is forbidened even under the $p$-doping
and the short plasmon wavelength also can't be observed in this case.

Through the comparation (Fig.9) between the results of the local long-wavelength plasmon branches and the one obtained by RPA,
we found that the long-wavelength result is consistent with the local RPA result in small-$\bf q$ limit (long-wavelength),
and it's also agree with the experiment results\cite{Chen J}.
For the RPA result of plasmon dispersion in
critical electric field $E_{\perp}=0.017$ eV, the undamped plasmon model vanishes due to the 
vanishing gap ($m_{D}^{{\rm min}}=0$) and the joint between the 2$A$ and 2$B$ regions.
On the contrary, if thr electric field is far away from te critical point (i.e., the band gap is large enough),
the whole plasmon model is in the undamped region.

\section{Conclusions}
We analytically investigate the interband and intraband behavior of the monolayer and bilayer silicene with nonzero chemical potential (finite band filling) and
thus electron-hole asymmetry and away from the half-filling.
The dynamical polarization of silicene as well as the other graphene-like hexagonal lattice system, is closely related to the interband and intraband transition
including the scattering of the charged impurity, the screened potential of impurity which contributed by the Friedel oscillation when $\mu> m_{D}^{{\rm max}}$,
and the plasmon dispersion or dielectric function of the doped silicene (with finite $\mu$).
The plasmon damping (into the electron-hole excitation) in this paper is mainly focus on the Landau damping,
but in experiments, it's also impactful to detect the damping of near-field signal\cite{Chen J} which is due to the circular two-dimension wave-vector
of the plasmon
damping as $\sim r^{-1/2}$ where $r$ is the distance from the surface.
The near-field signal here can also be observed by the HHG which with intensity $I\sim\sqrt{E_{\perp}}$.
In fact, there exist the self-energy correction (like the self-consistent hybridization function)
in the plasmon resonance, but the electron-hole excitations in the damping region cancel such correlation and yielding the conclusions agreed 
with the RPA results\cite{Olevano V}.
Here we need to note that both the electron-hole excitation with optical/ acoustic plasmon damping,
and the power-law Friedel oscillation which may be sinusoidal or cosine or the superposition of the both (like
the case in critical electric field) of the charged impurity screened potential or the spin/charge density discussed in this paper,
all requires the low-energy (low-frequency) and low-momentum,
and the well preserved spin structure (orientations) under the not-too-high temperature.
Except that, the local-field effect\cite{Pellegrino F M D} of the lattice structure together with the 
induced standard deviations are ignored in our calculations,
since they have negligible effects in our homogeneous model under the low-temperature and low-momentum regime.
The neglect of the local-field effect also results in the decreasing of the number of the plasmon branches
due to the suppression of the intreband transition and the optical plasmon branch\cite{Pellegrino F M D}.
The bilayer silicene, in contrast to the monolayer silicene or the normal double-layer system like the double quantum-well,
has a interlayer hopping which leads to the polarization-dependent band parameter.
The interlayer hopping due to the finite layer separetion results in the plasmon dispersion which is not simply $\sim\sqrt{{\bf q}}$,
but linear with ${\bf q}$ in weak damped case, except in the long-wavelength limit (see the classical plasmon model dispersion in Fig.8),
which is consistent with the results of the bilayer or multilayer graphene\cite{Yuan S,Gamayun O V}.
The observed linear (weakly damped) plasmon dispersion for the classical bilayer silicene is similar to the high-energy $\pi$-plasmon
or the double quantun well,
or the case of conducting substrate which with strong metallic screening in the bulk semiconductor,
while for the two-dimension dice lattice, the strong screening due to the flat band structure also leading to the linear-like plasmon dispersion
compared to the monolayer graphene but with a pressed point in the $\omega={\bf q}=\mu$ (see Ref.\cite{Malcolm J D}).
That also agree with the fact that the electron-hole continuum emerges in the double-layer system when the separation lower to the critical value and then the plasmon
model becomes damped.
Finally, our results can also be applied to the other low-energy Dirac models or the topological insulators.

\end{large}
\renewcommand\refname{References}

\clearpage
Fig.1
\begin{figure}[!ht]
   \centering
   \begin{center}
     \includegraphics*[width=0.8\linewidth]{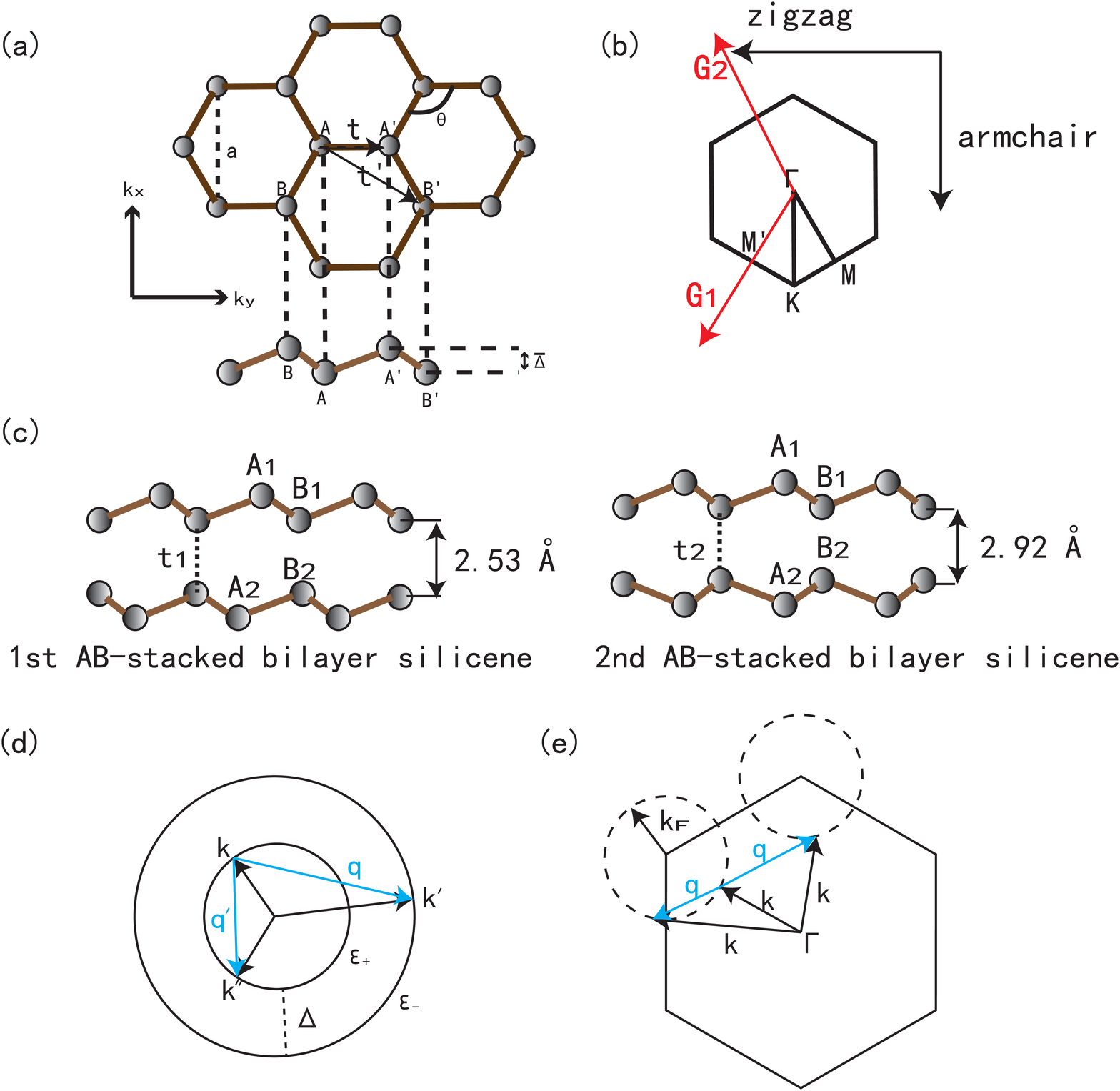}
\caption{(Color online) (a)Top view and side view of the silicene.
with four sites (sublattices) $A,B,A',B'$ in unit cell. 
 The bond-angle $\theta$ and the buckling distance $ m_{D}$ were marked.
The three dashed lines with $t,t',t''$ denotes the
nearest-, second nearest-, and third nearest-neighbor hopping, respectively.
The blue and green solid lines denotes the hopping in $r$ direction and $r'$ direction respectively,
where $r'$ contains the three hopping directions which goven by the phase $\phi$ and $r$ contains the three ones which not goven by the phase $\phi$.
(b) Brillouin zone (the $k$-space) with the high symmetry points.
The Red vector in the right panel is the reciprocal lattice vector ${\bf G}_{1}=(\frac{-2\sqrt{3}\pi}{3a},-\frac{2\pi}{a}),
{\bf G}_{2}=(\frac{-2\sqrt{3}\pi}{3a},\frac{2\pi}{a})$.
(d) the two kinds of the AB-stacked silicene: the first one
with the nearest layer distance as 5.2 \AA\ and intra-layer bond length 2.28 \AA\ and with bulked distance $\overline m_{D}=0.46$ \AA\ the same as the monolayer one,
the second one with the nearest layer distance as 2.46 \AA\ and intra-layer bond length 2.32 \AA\ and with lattice constant $a=3.88$,
and the bulked distance becomes $\overline m_{D}=0.64$ \AA\.
The interlayer hopping label in the figure are $t_{1}=2.025$ eV.
(d) Schematic of the scattering phase space for the interband scattering wave vector ${\bf q}$ and intraband scattering wave vector ${\bf q}'$.
The upper band $\epsilon_{+}$ and lower band $\epsilon_{-}$ and the band gap $ m_{D}$ are labeled in the figure.
(e) Schematic of the scattering in Brillouin zone for the intervalley scattering
wave vector ${\bf q}$ and intravalley scattering wave vector ${\bf q}'$.
The dash-circle is the Fermi surface, and the Fermi wave vector ${\bf k}_{F}$ is indicated which constitute the Fermi patchs,
and with the Fermi energy $E_{F}=\gamma {\bf k}_{F}$.
}
   \end{center}
\end{figure}
\clearpage
Fig.2
\begin{figure}[!ht]
   \centering
   \begin{center}
     \includegraphics*[width=0.8\linewidth]{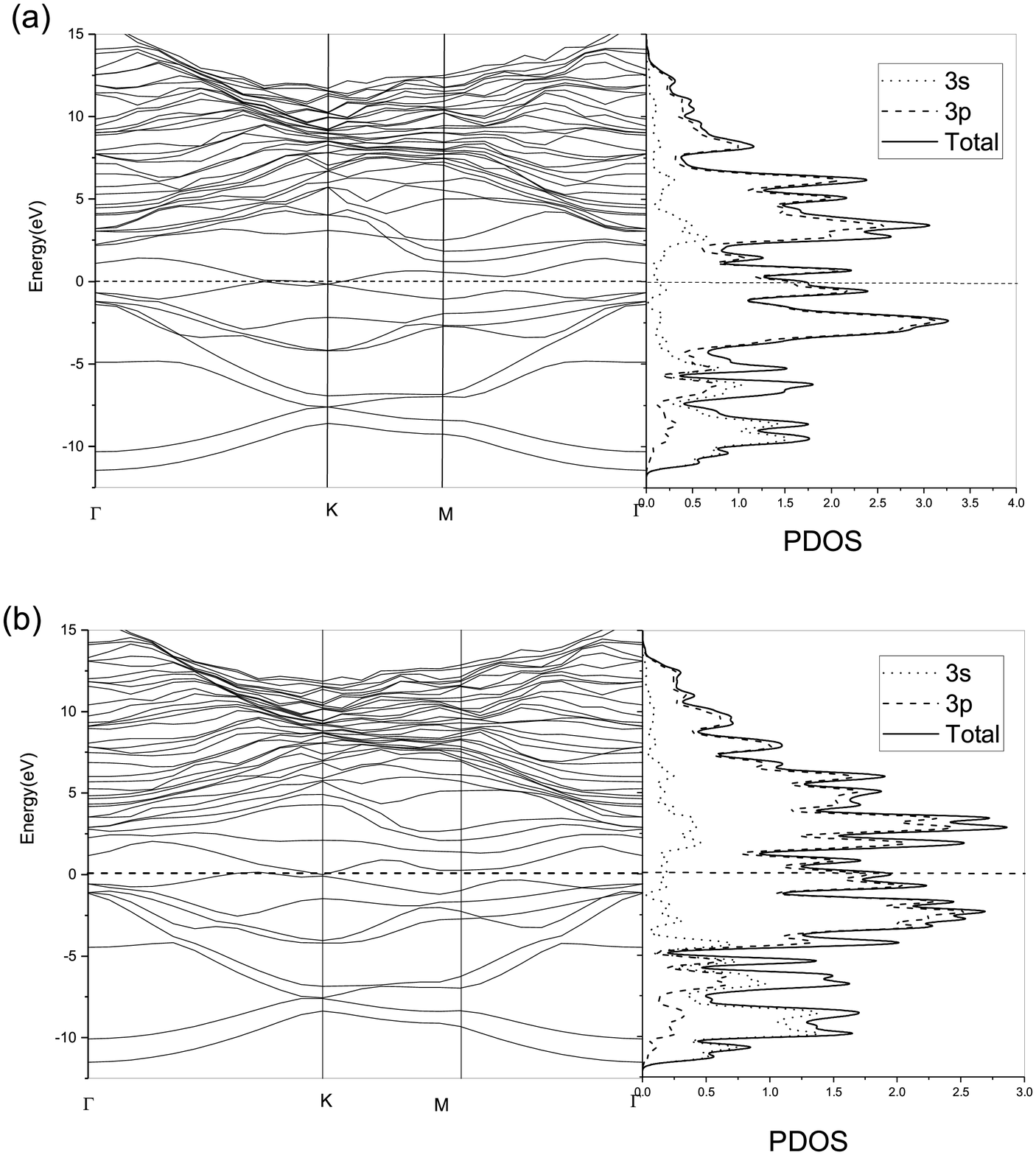}
\caption{Band structure of 1st AB-stacked bilayer silicene (a), and 2nd AB-stacked bilayer silicene (b) as well as their PDOS in the right side.
}
   \end{center}
\end{figure}

\clearpage
Fig.3
\begin{figure}[!ht]
   \centering
   \begin{center}
     \includegraphics*[width=0.8\linewidth]{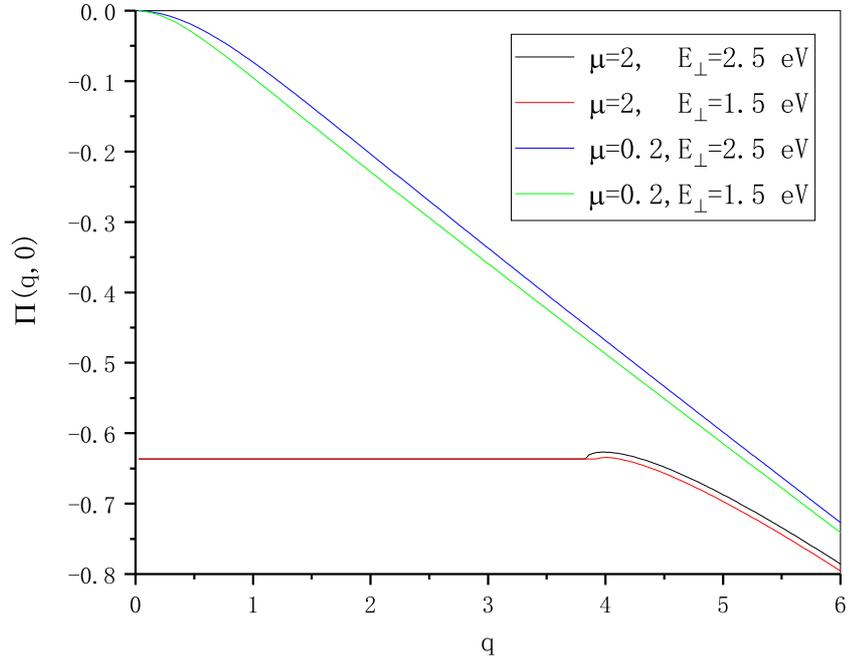}
\caption{(Color online) The static polarization $\Pi({\bf q},0)$ of silicene.
The black and red lines corresponds to the case of $\mu>m_{D}$ 
while the blue and green line corresponds to the case of $\mu<m_{D}$
(here $m_{D}=\lambda_{SOC}|\frac{\frac{\overline\Delta}{2}E_{\perp}}{\lambda_{SOC}}\pm 1|$ where $\pm$ sign
corresponds to the $m_{D}^{{\rm max}}$ and $m_{D}^{{\rm min}}$, respectively).
}
   \end{center}
\end{figure}

\clearpage
Fig.4
\begin{figure}[!ht]
\subfigure{
\begin{minipage}[t]{0.5\textwidth}
\centering
\includegraphics[width=0.9\linewidth]{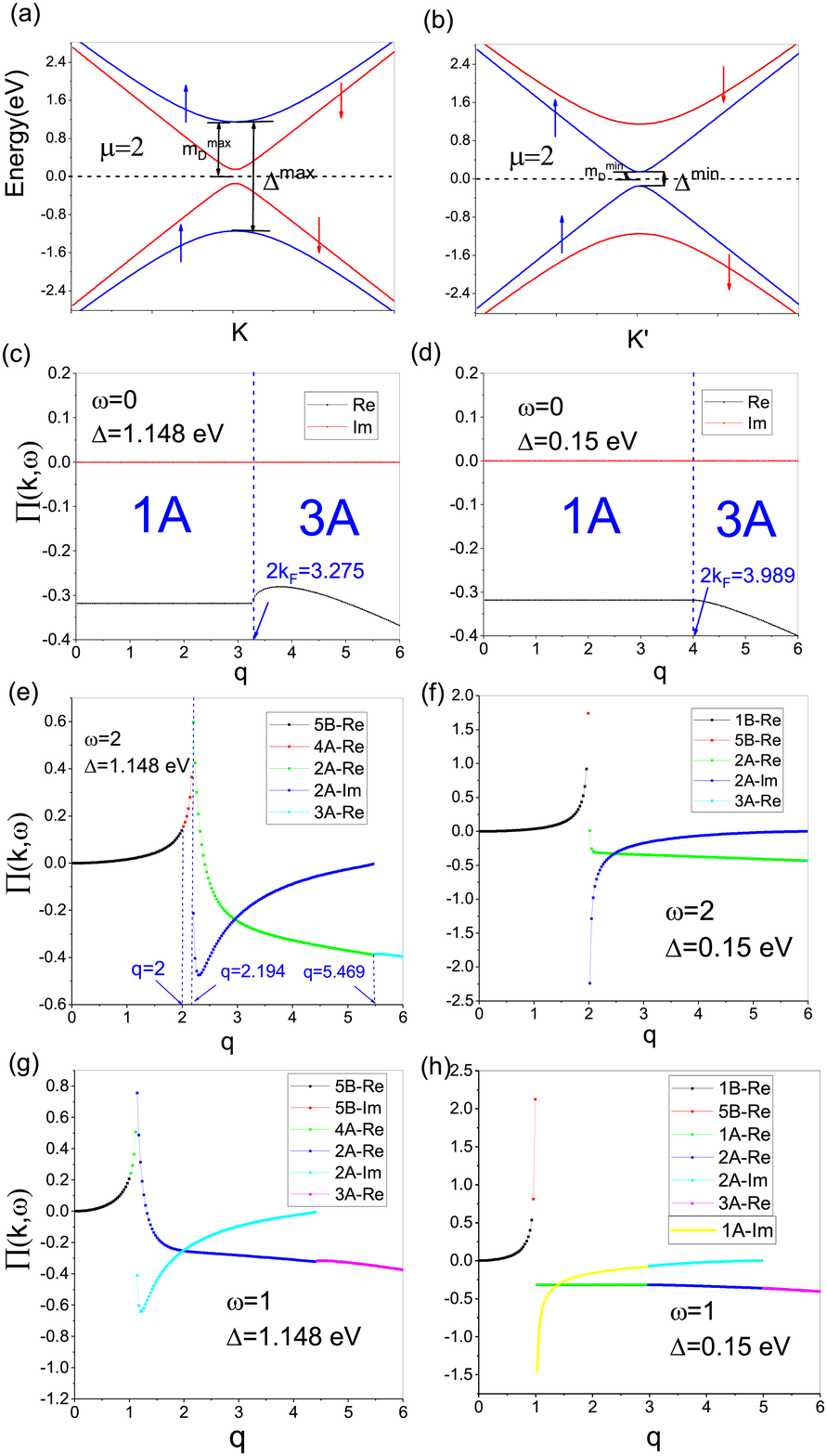}
\label{fig:side:a}
\end{minipage}
}\\
\subfigure{
\begin{minipage}[t]{0.5\textwidth}
\centering
\includegraphics[width=0.9\linewidth]{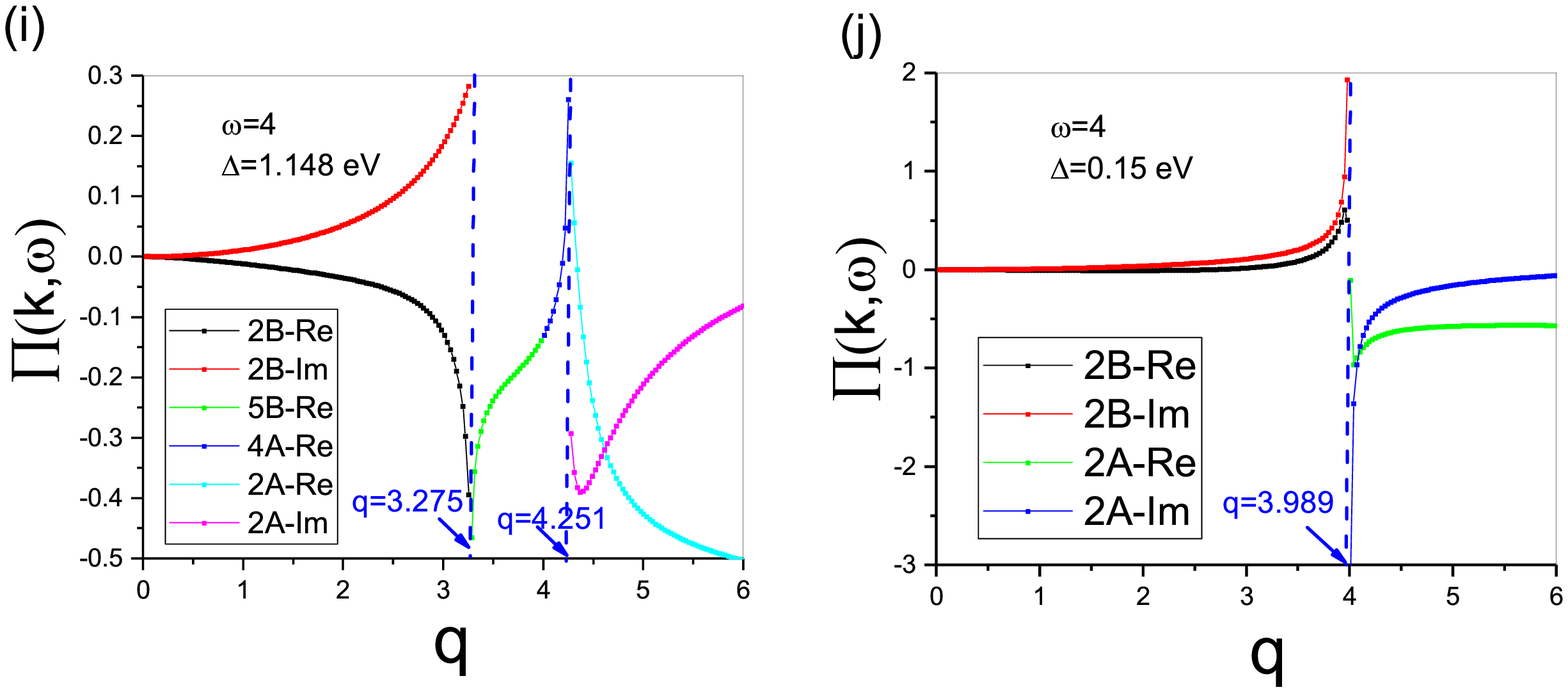}
\label{fig:side:b}
\end{minipage}
}
\caption{(Color online) 
Low-energy band structure of the monolayer silicene and the polarization function with finite $\mu$.
The on-site Hubbard interaction, Rashba-coupling, exchange field are setted as zero and the factors $\hbar$ and $v_{F}$ are setted as 1 in this figure and
the chemical potential is setted as $\mu=2$ here which is larger than the
Dirac-mass $ m_{D}=$1.148 eV in K valley and 0.15 eV in K' valley for the up-spin, i.e., $\mu> m_{D}^{{\rm max}}$
with the filled valence band and conduction band.
The maximum gap $\Delta^{{\rm max}}=2m_{D}^{{\rm max}}=2.296$ eV
is correponds to the maximum Fermi wave vector ${\bf k}_{F}^{{\rm max}}$, while the minimum gap correponds to $m_{D}^{{\rm min}}=0.15$ eV 
and the minimum Fermi wave vector ${\bf k}_{F}^{{\rm min}}$.
(a) The band structure of valley K and K' under the perpendicular electric field $E_{\perp}=1.321$ eV.
The bule line and red line correspond to the up-spin and down-spin, respectively.
In the case of $ m_{D}<\mu$,
the two regions labeled by the blue label A and B, respectively, have the distinct polarization 
(including the interband polarization and intraband polarization).
The singular point for the static dielectric function are labeled by 2${\bf k}_{F}$ in the (c)-(d) where the static polarization function
shows discontinuity in the first derivative,
and corresponds to ${\bf k}={\bf k}+{\bf q}={\bf k}_{F}$ with $\langle {\bf k},{\bf q}\rangle=\pi$\cite{Lin M F}.
The polarization for the frequency $\omega=2$, $\omega=1$, and $\omega=4$ are presented in (e)-(f), (g)-(h), and (i)-(j), respectively.
}
\end{figure}


\clearpage
Fig.5
\begin{figure}[!ht]
\subfigure{
\begin{minipage}[t]{0.45\textwidth}
\centering
\includegraphics[width=1.1\linewidth]{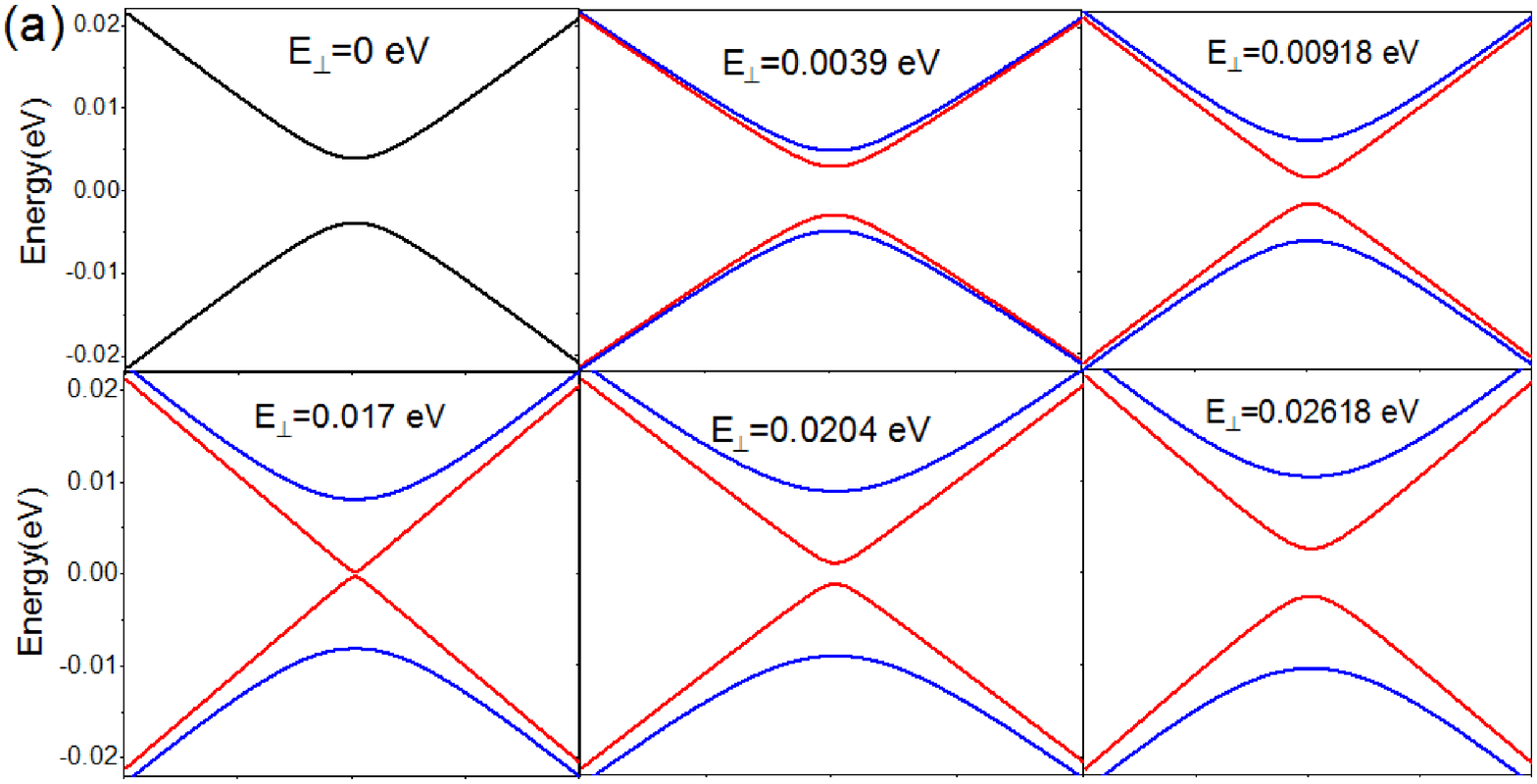}
\label{fig:side:a}
\end{minipage}
}\\
\subfigure{
\begin{minipage}[t]{0.5\textwidth}
\centering
\includegraphics[width=1.1\linewidth]{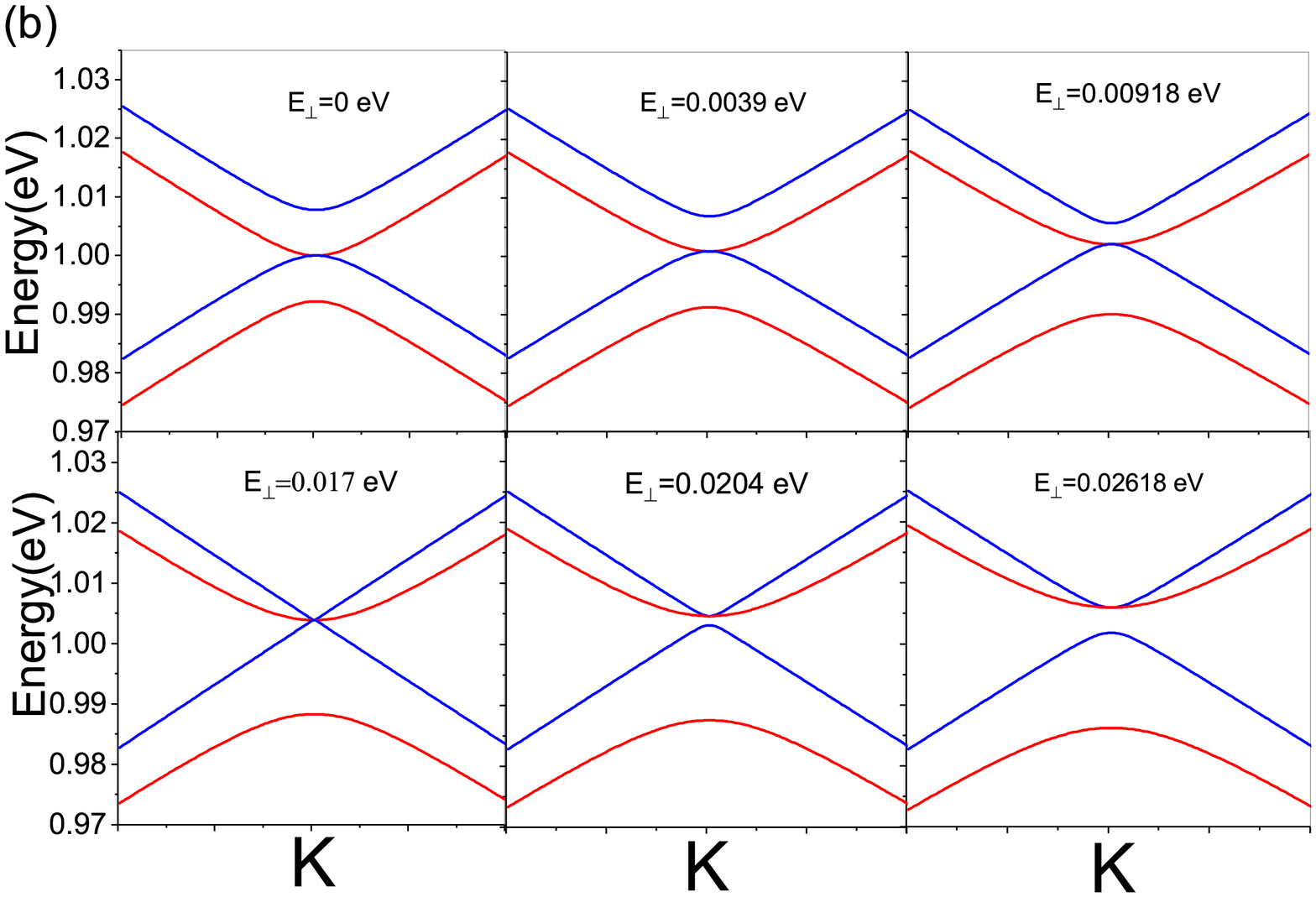}
\label{fig:side:b}
\end{minipage}
}\\
\subfigure{
\begin{minipage}[t]{0.5\textwidth}
\centering
\includegraphics[width=1.1\linewidth]{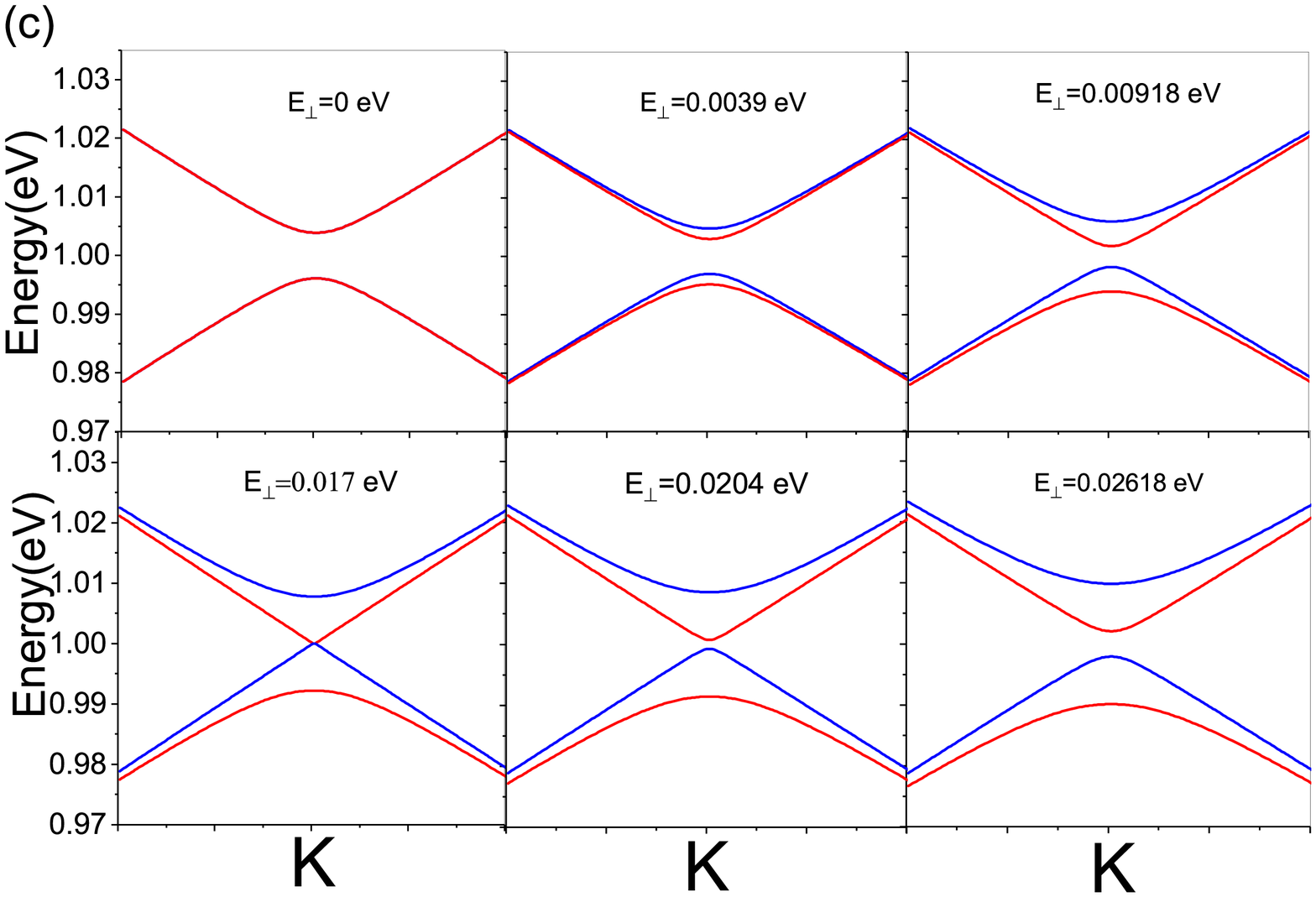}
\label{fig:side:b}
\end{minipage}
}
\caption{(Color online) Band gap evolution in valley K for the silicene nanoribbon under the effect of electric field and
effective SOC (a), additionally,
the electric-field-induced NN Rashba-coupling $R_{2}(E)$ are considered in (b),
and both the SOC, $R_{2}(E)$, exchange field $M$ which is setted as 3.9 meV here are all taken into account in (c).
The on-site interaction U is setted as zero for simplicity here, and the critical
electric field is arounds at 17 mev/A. The small NNN intrinsic Rashba-coupling is 0.7 meV.
The blue bands are correspond to the one with up-spin electrons while the red bands are correspond to the down-spin one.
We can see that there sequence of colors from top to bottom is different between (a) and (b)(c),
which also reveal the effects of the Rashba-coupling and exchange field.
}
\end{figure}

\clearpage
Fig.6
\begin{figure}[!ht]
\subfigure{
\begin{minipage}[t]{0.5\textwidth}
\centering
\includegraphics[width=1.2\linewidth]{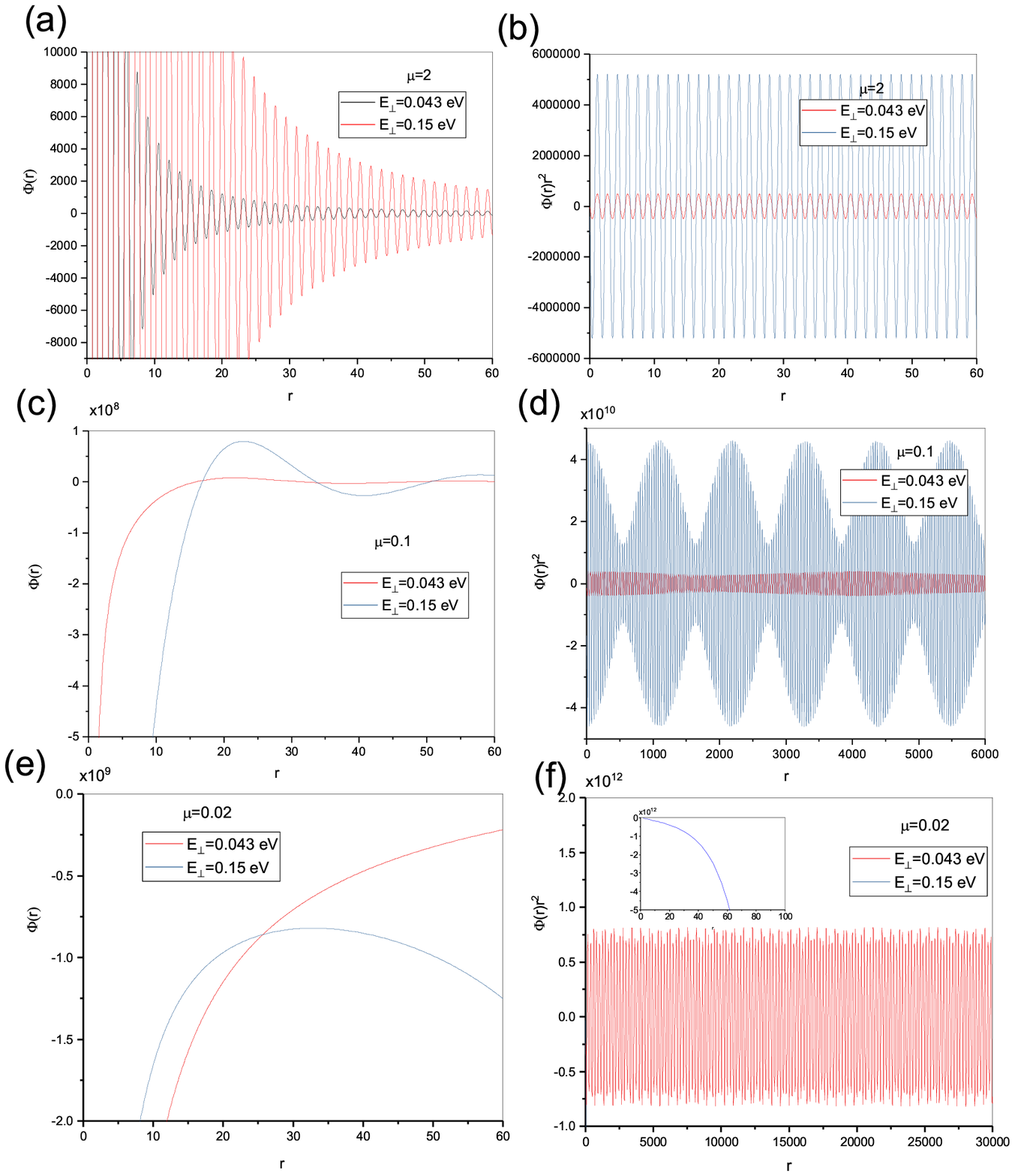}
\label{fig:side:a}
\end{minipage}
}\\
\subfigure{
\begin{minipage}[t]{0.5\textwidth}
\centering
\includegraphics[width=1.1\linewidth]{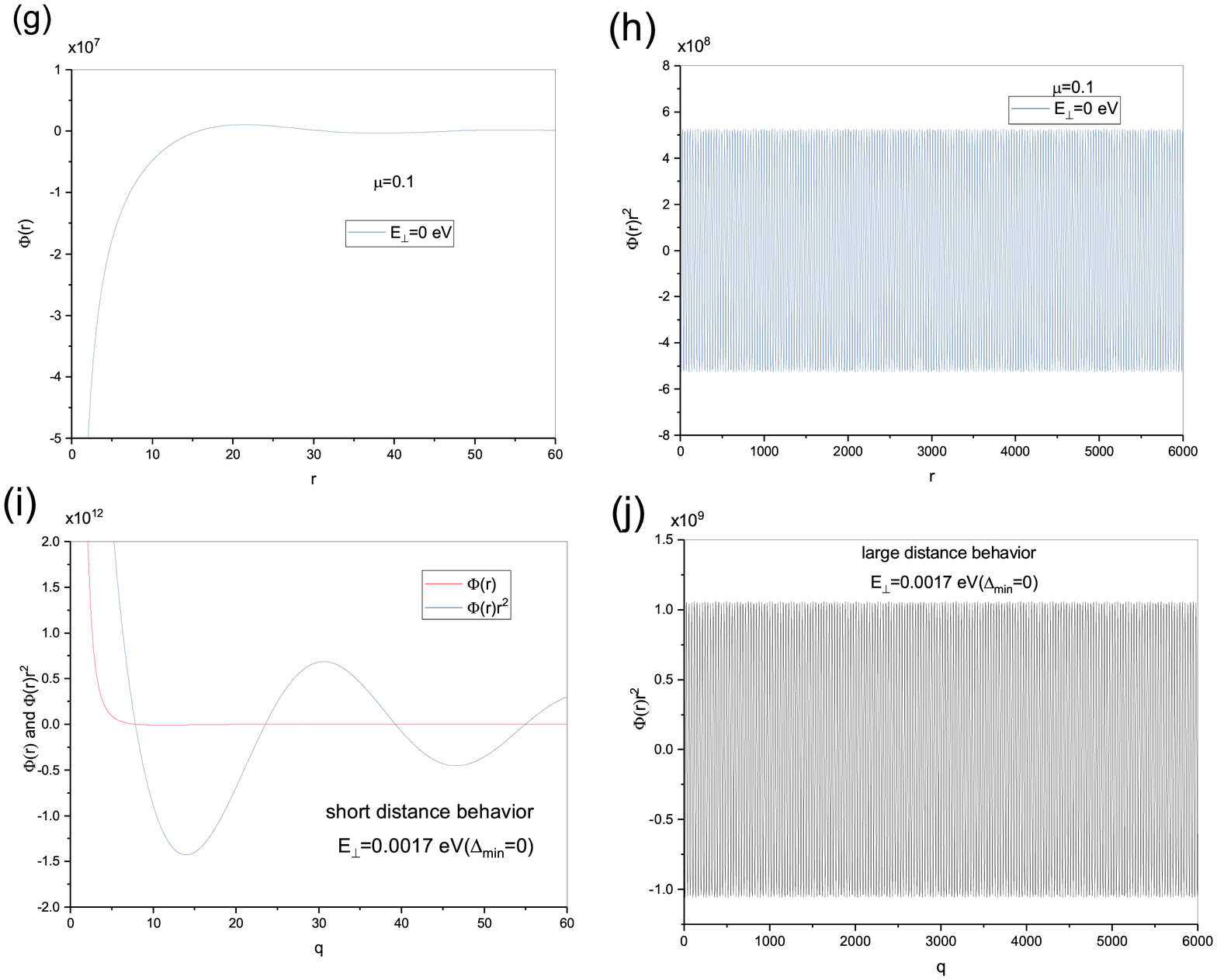}
\label{fig:side:b}
\end{minipage}
}
\caption{(Color online) 
Screened potential of the charged impurity at short distance (first column) 
and its asymptotic behavior ($r^{-2}$ decay (second column)) at large distance.
The impurity concentration is setted as $n_{{\rm imp}}=10^{12}$ cm$^{-2}$,
and we also set $e=1$,
$\hbar v_{F}=3.29$ eV, $\epsilon_{0}\epsilon=2.45$.
(a)-(b) are for chemical potential $\mu=2$, (c)-(d) for $\mu=0.1$, and (e)-(f) are for $\mu=0.02$.
In (a)-(f) we present both the results of under the electric field of $E_{\perp}=0.043$ eV and $E_{\perp}=0.15$ eV,
in (g)-(j), we present the results for $E_{\perp}=0.017$ eV and $E_{\perp}=0$ eV,
where we can easily see that the beatings in $\Phi(r)r^{2}$ are vanish.
}
\end{figure}
\clearpage
Fig.7
\begin{figure}[!ht]
   \centering
   \begin{center}
     \includegraphics*[width=0.8\linewidth]{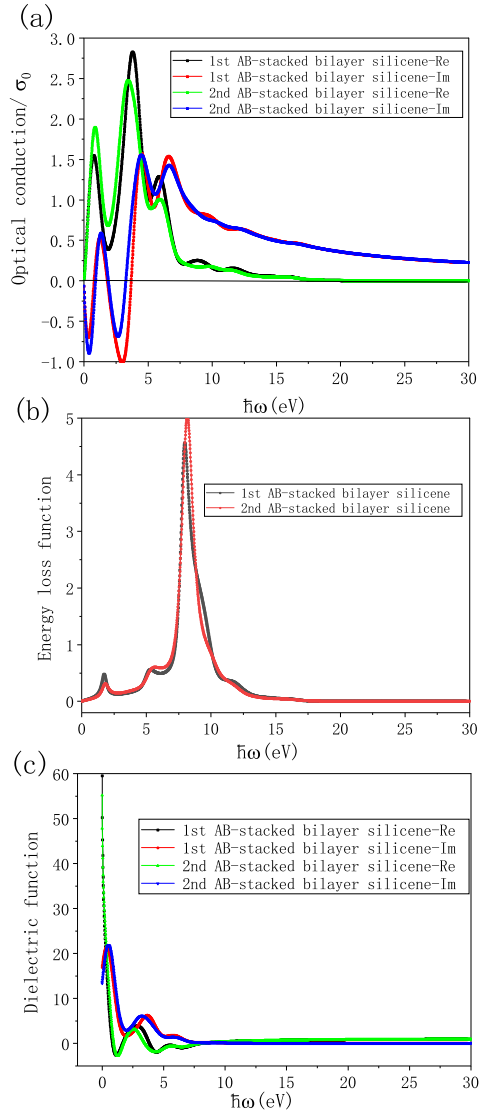}
\caption{(Color online) Optical conductivity in unit of universe ac constant conductivity $\sigma_{0}=e^{2}/(4\hbar)$ (a)
and energy loss function (b) and dielectric function (c) of the 1st AB-stacked bilayer silicene and 2nd AB-stacked bilayer silicene.
In (c), the plasmon pole can be seen easily and it's undamped due to the small ${\bf q}$ characteristic.
}
   \end{center}
\end{figure}
\clearpage
Fig.8
\begin{figure}[!ht]
   \centering
   \begin{center}
     \includegraphics*[width=0.8\linewidth]{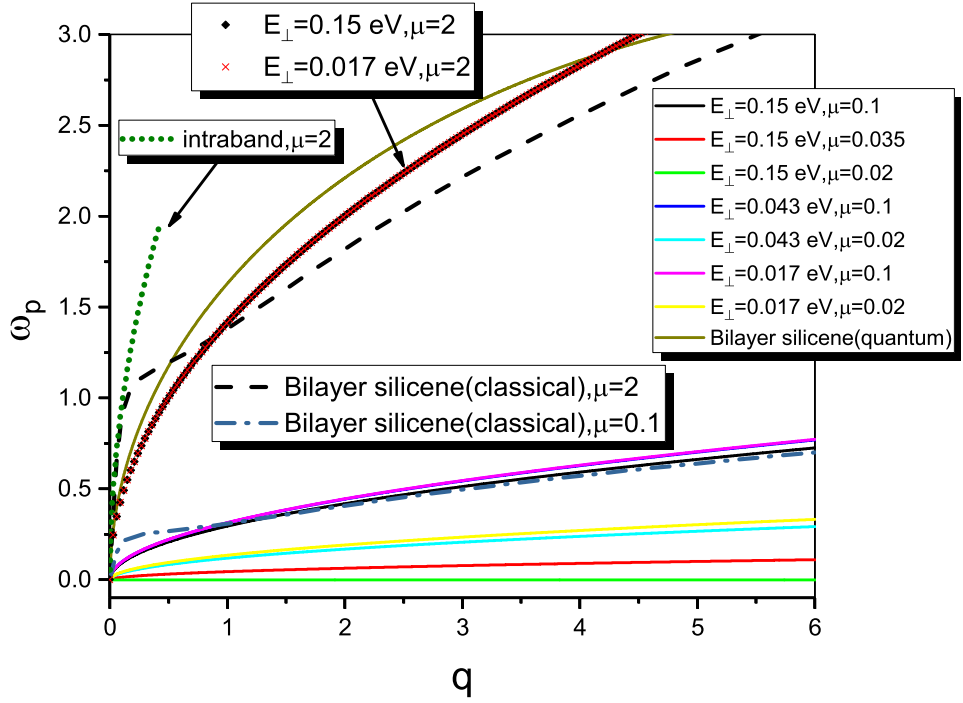}
\caption{(Color online) Plasmon frequency in long-wavelength case of the silicene which is electric field- (band gap) and chemical potential-dependent,
and the quantal bilayer silicene which is $r_{w}$- and density-dependent (obtained by Eq.(34)),
and the classical bilayer silicene.
The band parameter $\gamma$ is setted close to the interlayer hopping parameter 0.72, thus the Wigner-Seitz radius $r_{w}$ is estimated as 0.567
for bilayer silicene on SiO$_{2}$ substrate with the electron density $n=10^{12}$ cm$^{-2}$
(the $r_{w}$ would be as large as 11.34 for the freestanding bilayer silicene which is suspended and thus $\epsilon=1$).
For classical bilayer silicene, we only focus on the 1st AB-stacked bilayer silicene, i.e., interlayer distance $d=2.53$ \AA\,
and the interlayer vector is setted as $k_{z}=\pi/6$.
The dot-green line describes the contributions from the intraband transition 
where we set the temperature as $T=1$ K and $\mu=2$, and it's found very close to the result of the classical bilayer silicene in the region $\omega_{p}<1$.
The black rhombus and red crosses corresponds to the dispersion with $\mu=2$ under the electric field $E_{\perp}=0.15$ eV and 0.017 eV, respectively,
and we found that they are very close to each other which suggest that the electric field-dependence of the dispersion decrese 
with the increase of the chemicl potential.
}
   \end{center}
\end{figure}
\clearpage
Fig.9
\begin{figure}[!ht]
   \centering
   \begin{center}
     \includegraphics*[width=0.8\linewidth]{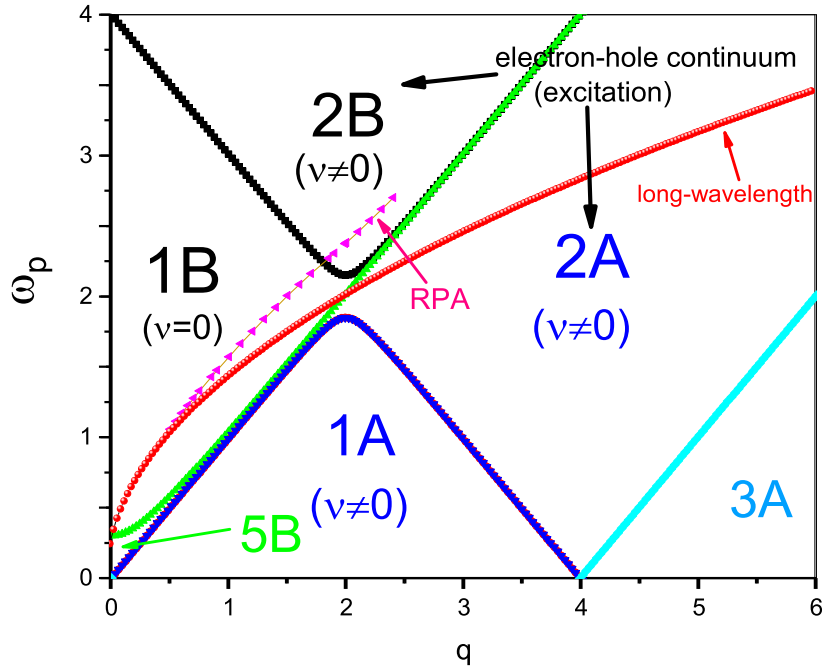}
\caption{(Color online) 
Plot of the regions with different characteristics of the polarizatio function which is
far away from the case of half-filling (which is simply devided by the line $\omega={\bf q}$), 
and the
numerical solution of the plasmon frequency in doped silicene ($\mu=2$) obtained by RPA,
${\rm Re}epsilon({\bf q},\omega_{p})=0$ in the weak damping region where the decay rate $\nu$ is ignored.
The chemical potential is setted as $\mu=2$, and the electric field is $E_{\perp}=0.67$ eV
which leads to the Dirac-mass: $ m_{D}^{{\rm max}}=0.1578$ eV, $ m_{D}^{{\rm min}}=0.15$ eV.
The region 2B is the interband single-particle excitation regime, while the $1A$ and $2A$ regimes are the intraband one.
The Wigner-Seitz radius is setted as $r_{w}=0.56$ here and the minimum Dirac-mass $ m_{D}^{{\rm min}}=0.15$ eV.
We also find that the shortest distance between $2B$ and $1A$ regions is roughly twice of the $ m_{D}^{{\rm min}}$ as 0.3 eV.
The interband (2$B$) and intraband ($1A$,$2A$) single-particle excitation regimes which with $\nu\neq 0$ are indicated.
The red circles corresponds to the long-wavelength result of the plasmon dispersion,
and the purple triangles corresponds to the RPA results obtained by solving the relation ${\rm Re}[\epsilon({\bf q},\omega_{p})]=0$.
}
   \end{center}
\end{figure}

\end{document}